# Discovery of Orbital-Selective Cooper Pairing in FeSe

P.O. Sprau[1,2†], A. Kostin[1,2†], A. Kreisel[3,4†], A. E. Böhmer[5], V. Taufour[5], P.C. Canfield[5,6], S. Mukherjee[7], P.J. Hirschfeld[8], B.M. Andersen[3] and J.C. Séamus Davis[1,2,9,10*]

1. LASSP, Department of Physics, Cornell University, Ithaca, NY 14853, USA.
2. CMPMS Department, Brookhaven National Laboratory, Upton, NY 11973, USA.
3. Niels Bohr Institute, University of Copenhagen, Juliane Maries Vej 30, DK 2100 Copenhagen, Denmark
4. Institut für Theoretische Physik, Universität Leipzig, D-04103 Leipzig, Germany
5. Ames Laboratory, U.S. Department of Energy, Ames, IA 50011, USA
6. Department of Physics and Astronomy, Iowa State University, Ames, IA 50011, USA.
7. Department of Physics, Binghamton University - SUNY, Binghamton, USA.
8. Department of Physics, University of Florida, Gainesville, Florida32611, USA
9. School of Physics and Astronomy, University of St. Andrews, Fife KY16 9SS, Scotland.
10. Tyndall National Institute, University College Cork, Cork T12R5C, Ireland.
† These authors contributed equally to this work.
* Correspondence to: jcseamusdavis@gmail.com

**1** **The superconductor FeSe is of intense interest thanks to its unusual non-magnetic nematic state and potential for high temperature superconductivity. But its Cooper pairing mechanism has not been determined. Here we use Bogoliubov quasiparticle interference imaging to determine the Fermi surface geometry of the bands surrounding the $\Gamma=(0,0)$ and $X=(\pi/a_{Fe}, 0)$ points of FeSe, and to measure the corresponding superconducting energy gaps. We show that both gaps are extremely anisotropic but nodeless, and exhibit gap maxima oriented orthogonally in momentum space. Moreover, by implementing a novel technique we demonstrate that these gaps have opposite sign with respect to each other. This complex gap configuration reveals the existence of orbital-selective Cooper pairing which, in FeSe, is based preferentially on electrons from the $d_{yz}$ orbitals of the iron atoms.**

**2** The high temperature superconductivity in iron-based superconductors is typically most robust where coexisting antiferromagnetic and nematic ordered states are suppressed by doping or pressure (*1-3*). However, FeSe appears distinctive for several reasons: (i) Although strongly nematic, it does not form an ordered magnetic state and is instead hypothesized to be a quantum paramagnet (*4-6*); (ii) it exhibits evidence for orbital selectivity (*7,8*) of band structure characteristics (*9-12*); (iii) a monolayer of FeSe grown upon a SrTiO$_3$ substrate produces the highest T$_c$ of all iron-based superconductors (*13-16*). It is therefore essential to understand the electronic structure and superconductivity of FeSe at a microscopic level; however, the Cooper pairing mechanism of FeSe is unknown. A quantitative determination of the momentum space ($\vec{k}$-space) structure and relative sign of the superconducting energy gaps $\Delta_i(\vec{k})$ on each electronic band $E_i(\vec{k})$ is necessary to identify this mechanism. So far, this has not been achieved because of the minute Fermi-surface pockets as well as the highly anisotropic $\Delta_i(\vec{k})$ requiring energy resolution δE<100μeV. Here we use sub-kelvin Bogoliubov quasiparticle interference imaging (BQPI) (*17-19*), an established technique for high-precision multiband $\Delta_i(\vec{k})$ determination (*20,21,22*), to measure the detailed structure of the energy gaps in FeSe.

**3**      In the orthorhombic phase below $T_S \cong 90K$, FeSe has a crystal unit cell with conventional lattice parameters $a$=5.31 Å, $b$=5.33 Å and $c$=5.48 Å. Here we parameterize the Fe-plane of the same lattice using the two inequivalent Fe-Fe distances $a_{Fe}$=2.665 Å and $b_{Fe}$=2.655 Å in the orthorhombic/nematic phase (Fig. 1A, section I of supplementary materials (SM)); we define the $x$-axis ($y$-axis) to always be parallel to the orthorhombic $a_{Fe}$-axis ($b_{Fe}$-axis), so that our x/y coordinate system rotates when a twin boundary is crossed. The FeSe Fermi surface (FS) is postulated to consist of three bands α, ε and δ (shown for $k_z$=0 in Fig. 1B), and may be parameterized accurately using a tight-binding model (*23,24*) that is fit simultaneously to several types of experimental observations (sections II and III of SM). Surrounding the Γ=(0,0) point is an ellipsoidal hole-like α-band, whose FS $\vec{k}_\alpha(E=0)$ has its major axis aligned to the orthorhombic $b_{Fe}$-axis; surrounding the X=($\pi/a_{Fe}$,0) point is the electron-like ε-band whose "bowtie" FS $\vec{k}_\varepsilon(E=0)$ has its major axis aligned to the orthorhombic $a_{Fe}$-axis. At the Y=(0,$\pi/b_{Fe}$) point, a δ-band FS should also exist but has not been detectable by spectroscopic techniques. In this picture, the $d_{yz}$ orbital content of the α-band Fermi surface has its maximum value along the $x$-axis (green in Fig. 1B) whereas its $d_{xz}$ orbital content peaks along the $y$-axis (red in Fig. 1B). Conversely, the $d_{yz}$ orbital content of the ε-band FS is maximum along the $y$-axis (green in Fig. 1B), and its $d_{xy}$ orbital content reaches its highest point along the $x$-axis (blue in Fig. 1B); (Refs. *23,24,25* and section II of SM). These α-band and ε-band FS pockets (Fig. 1B) exhibit maximal simultaneous consistency with FS geometry from angle resolved photoemission (ARPES) (*25,26*), quantum oscillations (QO) (*27,28*), and BQPI as discussed below (section III of SM).

**4**     A fundamental issue in iron-based superconductivity research is whether conduction electrons are weakly or strongly correlated, and the consequences thereof for enhancing the superconductivity. The situation is complex because multiple Fe orbitals (e.g. $d_{xz}$, $d_{yz}$, $d_{xy}$) are involved. One limit of theoretical consideration is an uncorrelated metallic state where 'nesting' features of the FS geometry generate antiferromagnetic spin-fluctuations which then mediate Cooper pairing and superconductivity (*2*). By contrast, the ordered magnetic states of these same materials are often modeled using frustrated multi-orbital Heisenberg ($J_1$-$J_2$) models in which electrons are essentially localized, with the metallicity and spin-fluctuation-mediated-superconductivity appearing upon doping this magnetic insulator (*5*). Intermediate between the two is the Hund's metal viewpoint (*7*) in which strong Hund's coupling, while aligning the Fe spins, also suppresses the inter-orbital charge fluctuations. This generates orbital decoupling in the electronic structure which allows 'orbital selectivity' to occur in the effects of correlations (*7,8*). In theory, the result can be Mott-localized states associated with one orbital coexisting with delocalized quasiparticle states associated with others. Under such circumstances, the pairing itself might become orbital-selective (*29,30*) meaning that the electrons of predominantly one specific orbital character bind to form the Cooper pairs of the superconductor. If this occurs, the superconducting energy gaps should become highly anisotropic (*29,30),* being large only for those FS regions where a specific orbital character dominates. Such phenomena have remained largely unexplored because orbital-selective Cooper pairing has never been detected in any material.

**5**     To search for such pairing in FeSe, we apply BQPI imaging of impurity-scattered quasiparticles that interfere quantum-mechanically to produce characteristic modulations of the density-of-states, $N(\vec{r}, E)$ surrounding each impurity atom. When a $\vec{k}$-space energy gap $\Delta_i(\vec{k})$ is anisotropic, the Bogoliubov quasiparticle dispersion $E_i(\vec{k})$ will exhibit closed constant-energy-contours (CEC) which are roughly 'banana-shaped' and surround FS points where $\Delta_i(\vec{k})$ is minimum (*20-22*). Then, at a given energy E, the locus of the 'banana tips' can be determined because the maximum intensity BQPI modulations occur at wavevectors $\vec{q}_j(E)$ connecting the tips, thanks to their high joint density of states (JDOS) for scattering interference. Both the superconductor's Cooper-pairing energy gap $\Delta_i(\vec{k})$ and the FS on each band are then determined directly (*20-22*) by geometrically inverting the measured BQPI wavevector set $q_j(E)$ in the energy range $\Delta_i^{min}$<E<$\Delta_i^{max}$. As these techniques can be implemented at temperatures T $\leq$ 300$mK$, the $\Delta_i(\vec{k})$ on multiple bands can be measured with energy resolution $\delta E \approx$ 75$\mu eV$ (*21,22*), a precision unachievable by any other approach.

**6** However, no BQPI measurements have been reported for bulk FeSe, although photoemission data for the equivalent of the α-band do exist for a related compound Fe(Se,S) (*31*). For guidance, we first consider a pedagogical model but note that Fermi surfaces and energy-gap structures derived using BQPI imaging do not depend on any particular model (*20-22*). Given the *α*-band FS (fine dashed grey contour in Fig. 1C) supporting an anisotropic $\Delta_\alpha(\vec{k})$ that has C$_2$ symmetry (*31,32*), the contours-of-constant energy (CEC) would be as shown by the fine colored curves, with quasiparticle energy increasing as indicated by the color code. The tips of each Bogoliubov CEC 'banana' are then indicated by colored dots similarly representing increasing energy; we expect that a triplet of inequivalent BQPI wavevectors $\vec{q}_i^\alpha(E)$ i=1-3 should exist (black arrows in Fig. 1C). The anticipated energy dependence of the $\vec{q}_i^\alpha(E)$ is shown schematically in Fig. 1E using the same color code as for banana-tips in Fig. 1C. For each energy $\Delta_i^{min} < E < \Delta_i^{max}$ the positions of the four CEC bananas-tips $(\pm k_x(E), \pm k_y(E))_\alpha$ can be determined by inverting

$$\vec{q}_1^\alpha = (0, 2k_y) \qquad (1)$$

$$\vec{q}_3^\alpha = (2k_x, 0) \qquad (2)$$

$$\vec{q}_2^\alpha = (2k_x, 2k_y) \qquad (3)$$

If a C$_2$-symmetric energy gap $\Delta_\delta(\vec{k})$ existed on the δ-band surrounding Y=(0,π/b$_{Fe}$) it might be expected to behave very comparably. A similar analysis (Fig. 1, D and F) applies to the "bowtie" $\varepsilon$-band FS surrounding X=($\pi$/a$_{Fe}$,0) (grey contour in Fig. 1D) with the anticipated energy dependence of the $\vec{q}_i^{\varepsilon}(E)$ shown schematically in Fig. 1F.

**7** To measure the FS and the superconducting gap structure and sign, we image differential tunneling conductance $dI/dV(\vec{r}, eV) \equiv g(\vec{r}, E)$ at T=280mK both as a function of location $\vec{r}$ and electron energy $E$. As the FS pockets are so miniscule in area (Fig. 1B), the expected range of dispersive intraband BQPI wavevectors is very limited $0<|\vec{q}_i^{\alpha,\varepsilon}(E)| < 0.25(\frac{2\pi}{a_{Fe}})$, whereas the interband BQPI necessitates resolving wavevectors $\geq \frac{\pi}{a_{Fe}}$. To achieve the $\vec{q}$-space resolution $|\delta q_i^{\alpha,\varepsilon}| \leq 0.01\left(\frac{2\pi}{a_{Fe}}\right)$ required to discriminate the energy evolution of BQPI on both $\alpha$-band and $\varepsilon$-band necessitates high-precision $g(\vec{r}, E)$ imaging in very large fields of view. We typically use between 60X60 nm² and 90X90 nm² (section IV of SM). Local maxima of $|g(\vec{q}, E)|$, the amplitude Fourier transform of $g(\vec{r}, E)$, are then used to determine the characteristic wavevectors $\vec{q}_i^{\alpha}(E)$ and $\vec{q}_i^{\varepsilon}(E)$ of dispersive modulations of BQPI. Figure 2A shows a typical example of measured $g(\vec{r}, E)$ with its $|g(\vec{q}, E)|$ in Fig. 2C. Here, by using a low resolution STM-tip, we predominantly detect the BQPI signal corresponding to the $\alpha$-band surrounding Γ=(0,0) (a complete data set is shown in movie S1 and section IV of SM). The evolution of the BQPI triplet $\vec{q}_i^{\alpha}(E)$ (black crosses in Fig. 2C) in the range 2.3meV>|E|>0.8meV at 280mK is plotted in Fig. 2E. Analogous images for the $\varepsilon$-band obtained using very

high spatial resolution tips sensitive to states at high $\vec{k}$ are shown in Figs. 2, B, D, and F (a complete data set is shown in movie S2 and section IV of SM). Because both $\vec{q}_2^{\alpha}(E)$ and $\vec{q}_2^{\varepsilon}(E)$ evolve to finite wavevectors $2\vec{k}_F^{\alpha}$ and $2\vec{k}_F^{\varepsilon}$ respectively as $E \to 0$ (Fig. 2, E and F), FeSe superconductivity is in the BCS limit and not near the Bose-Einstein condensation limit where BQPI wavevectors must evolve to 0 as $E \to 0$. From the conventional $N(E) \equiv dI/dV(E)$ density-of-states spectrum at T=280mK (Fig. 2, G and H) we find that the maximum gap on any band is $\Delta_{\alpha}^{max} = 2.3$meV whereas another coherence peak occurs at the gap maximum of a second band at $\Delta_{\varepsilon}^{max} = 1.5$meV. The maximum gaps were assigned to each band based on the energy evolution of BQPI to the energy limit E → 2.3meV for the α-band and E → 1.5meV for the ε-band. Finally, because no conductance is detected in the energy region $E \lesssim 150\mu eV$, $\Delta^{min} \gtrsim 150\mu eV$ for all bands.

**8**     The FS for both the α- and ε-bands is next determined using the fact that the $\vec{k}$-space loci of CEC 'banana-tips' from both $\Delta_{\alpha}(\vec{k})$ and $\Delta_{\varepsilon}(\vec{k})$ follow the FS of each band (Fig. 1 and Refs *20-22*). The measured evolution of the BQPI wavevector triplets $\vec{q}_i^{\alpha}(E)$ and $\vec{q}_i^{\varepsilon}(E)$ is plotted in Fig. 2, E and F. Figures 3, A and B show the two Fermi surfaces of the α- and ε-bands of FeSe determined from BQPI (section V of SM) using blue dots plus errors bars for each measured point and blue curves for the FS. The area of the Fermi surfaces extracted by BQPI is consistent with that at $k_z = 0$ (see section III of SM and Fig. S5). Next, we plot schematically the measured magnitude of the energy gap $|\Delta_\alpha(\vec{k})|$ on the $\alpha$-band in Fig. 3A, and the measured magnitude $|\Delta_\varepsilon(\vec{k})|$ on the $\varepsilon$-band in Fig. 3B, where in both cases we use the width of the gray shaded region to indicate $|\Delta(\vec{k})|$ and include values of extrema of any energy gap from N(E). Although exhibiting extraordinarily anisotropic ($\Delta_\alpha^{max}/\Delta_\alpha^{min} \gtrsim 15$) C2-symmetric energy-gap structures, FeSe remains a fully gapped or nodeless (*33-37*) superconductor with gap minima $\Delta_{\alpha,\beta}^{min} \gtrsim 150\mu eV$ .

**9** One of the key characteristics of iron-based superconductors is whether the energy gaps on different bands have opposite signs (*2,3*). For FeSe this situation should be designated $\pm$ because the more conventional designation $s_\pm$ (*2,3*) is rendered inappropriate by orthorhombic crystal/band-structure symmetry. One technique for measuring $\pm$ pairing symmetry is to detect the enhancement in amplitude of $g(\vec{q}, E)$ at specific BQPI wavevectors when a magnetic field is applied; this was proposed to occur because field-induced scattering results in amplified quasiparticle interference between regions of $\vec{k}$-space with same-sign energy gaps (*38*). In Fe(Se,Te) this approach has yielded field-induced QPI intensity reduction for wavevectors linking the electron and hole pockets, indicative of $\pm$ pairing symmetry (*39*). Yet, there are reservations about this interpretation (*40*) because: (i) a subset of wavevectors where the Fe(Se,Te) field-induced alternations are reported occur at Bragg points of the reciprocal-lattice and, (ii) a microscopic explanation for the field-induced reductions is absent. To address these issues, another BQPI technique designed to determine ± pairing symmetry has been proposed (*40*). It is based on conventional (non-magnetic) impurity scattering and the realization that the particle-hole symmetry of interband scattering interference patterns depends on the relative sign of the energy-gaps on those bands (*40*). As a result, the energy-symmetrized $\rho_+(\vec{q}, E)$ and energy-antisymmetrized $\rho_-(\vec{q}, E)$ phase-resolved Bogoliubov scattering interference amplitudes

$$\rho_\pm(\vec{q}, E) = Re\{g(\vec{q}, +E)\} \pm Re\{g(\vec{q}, -E)\} \qquad (4)$$

have, at the $\vec{q}$ for interband scattering, distinct properties depending on the relative sign of the two gaps. Importantly, this approach while not requiring variable temperature measurements requires phase-resolved imaging of BQPI in order to reliably discriminate $Re\{g(\vec{q}, E)\}$ from $Im\{g(\vec{q}, E)\}$. Moreover, the (anti)symmetrized functions $\rho_\pm(\vec{q}, E)$ must be integrated over a particular $\vec{q}$-space region. Specifically, we focus on

$$\rho_-(E) = \sum_{\delta\vec{q}} \rho_-(\vec{p}_1 + \delta\vec{q}, E) \qquad (5)$$

with radius $\delta q$ confining $\vec{q}$-space to interband scattering processes between two distinct energy gaps (Fig. 3C). Given our quantitative knowledge of the FS and energy gaps of FeSe (Fig. 3A,B), the $\rho_-(\vec{q}, E)$ can be predicted specifically for this material with the result shown as solid black curve in Fig. 3F for the FeSe gaps $\Delta_\alpha(\vec{k}), \Delta_\varepsilon(\vec{k})$ with $\pm$ pairing symmetry ( section VI of SM).

**10** Experimentally, the challenge is then to achieve phase-resolved imaging of BQPI surrounding a single impurity atom in FeSe, such as an Fe site vacancy (41,42). Therefore, we measure $g(\vec{r}, E)$ around individual impurity sites, each in a ~ 6.5x6.5 nm² field of view (see e.g. Fig. 3D) and then map the $g(\vec{q}, E)$ data onto a perfectly periodic atomic lattice. The $\vec{r}$-space origin of this lattice is then set at the impurity site (Fig. 3D) and $\rho_-(\vec{q}, E) = Re\{g(\vec{q}, +E)\} - Re\{g(\vec{q}, -E)\}$ is measured. Figure 3E is a typical example of $\rho_-(\vec{q}, E)$ (section VII of SM). Finally, the $\rho_-(E)$ is determined from Eq. 5 with the integration radius $\delta q$ chosen to capture only intensity related to scattering between the $\Delta_\alpha(\vec{k})$ and $\Delta_\varepsilon(\vec{k})$ inside the black circle in Fig. 3E. The resulting $\rho_-(E)$ is shown as black dots in Fig. 3F. Comparison of this measured $\rho_-(E)$ to its predicted form for the FeSe gaps $\Delta_\alpha(\vec{k}), \Delta_\varepsilon(\vec{k})$ with $\pm$ symmetry (solid black curve), shows them to be in good agreement especially in that $\rho_-(E)$ for $\pm$ pairing symmetry does not cross zero within the range of energy gaps. Thus, within the framework of Ref. 40, these data demonstrate that the sign of $\Delta_\alpha(\vec{k})$ is opposite to that of $\Delta_\varepsilon(\vec{k})$.

**11** Figures 4A,B summarize the key results of our study: the measured values of $\Delta_\alpha(\vec{k})$, $\Delta_\varepsilon(\vec{k})$ are both extremely anisotropic but nodeless, each having C$_2$- symmetry with deep minima that are aligned along orthogonal crystal axes. Recalling that our *x*-axis is defined to always be the orthorhombic $a_{Fe}$-axis, we have found these results to be equally true in both nematic domains. Such a gap structure is highly divergent from conventional spin fluctuation pairing theory (23) which yields a weak almost isotropic gap on the α-band and a strong gap on the ε–band but with an anisotropy of opposite $\vec{k}$-space orientation to that of the experimental data (section VIII of SM). Remarkably, however, orbital-selective pairing concentrated in the *dyz* channel can provide an explanation for the observed $\Delta_\alpha(\vec{k})$ and $\Delta_\varepsilon(\vec{k})$. Figure 4B shows our measured angular dependence of $\Delta_\alpha(\vec{k})$ about Γ=(0,0) and the equivalent for $\Delta_\varepsilon(\vec{k})$ about X=(π/$a_{Fe}$,0). For a *dyz* orbital-selective pairing interaction peaked at wavevector $\vec{q}$=(π/$a_{Fe}$,0), the predicted angular dependence of $\Delta_\alpha(\vec{k})$ and $\Delta_\varepsilon(\vec{k})$ is shown in Fig. 4C (section VIII of SM) and its comparison to the measured $\Delta_\alpha(\vec{k}), \Delta_\varepsilon(\vec{k})$ in Fig. 4B indicates the existence of orbital-selective Cooper pairing in FeSe.

**12** Microscopically, such orbital-selective Cooper pairing may arise from differences in correlation-strength for electrons with different orbital character. For example, correlations sufficient to generate incoherence for states with predominantly *dxy* orbital character (*11,12*) would suppress their pairing within an itinerant picture. Moreover, superconducting FeSe must exhibit distinct quasiparticle weights at the FS for states with $d_{xz}$ and $d_{yz}$ orbital character because of the nematic state (15,16). Under such circumstances, a Cooper pairing interaction focused at wavevector $\vec{q}=(\pi/a_{Fe},0)$ and forming spin-singlets from electrons predominantly with $d_{yz}$ orbital character can be modelled by enhancing the strength of spin-fluctuation pairing interaction for $d_{yz}$ orbital-character electrons relative to that of $d_{xz}$, while fully suppressing it for those with $d_{xy}$ orbital-character (Ref. 23 and section VIII of SM for details). Here, the orbital selectivity of pairing arises from quasiparticle weights in the various channels of itinerant spin-fluctuation pairing theory, which are hypothesized to be very different owing to orbital-selective correlations (section VIII of SM). Such a model could explain why the δ-band, predominantly associated with the $d_{xy}$ orbital, has weak visibility by ARPES (*11,12*) and BQPI (Figs. 2,3), and could also account for a low energy spin-susceptibility that is dominant at $\vec{q}=(\pi/a_{Fe},0)$ consistent with inelastic neutron scattering data (*43*). By projecting this form of orbital-selective pairing interaction onto the Fermi surfaces of FeSe (Fig. 3), the gap functions can be predicted by solving the linearized gap equation (Ref. 23 and section VIII of SM). The resulting predicted $\Delta_\alpha(\vec{k})$ and $\Delta_\varepsilon(\vec{k})$ (solid curves Fig. 4C) are quantitatively consistent with the extremely anisotropic structure and sign reversal of the measured gap functions (Fig. 4A,B). Moreover, as the magnitudes of $\Delta_\alpha(\vec{k})$ and $\Delta_\varepsilon(\vec{k})$ (solid curves Figs. 4B) track the strength of $d_{yz}$ orbital character on both bands (dashed curves Fig. 4C; Ref. *23,24,25;* section II of SM), the influence of orbital-selectivity on the Cooper pairing is manifest directly.

Overall, these data reveal a unique new form of correlated superconductivity based on orbital-selective Cooper pairing of electrons which, for FeSe, are predominantly from the $d_{yz}$ orbitals of Fe atoms. Such orbital selectivity may be pivotal to understanding the microscopic interplay of the quantum paramagnetism, nematicity and high temperature superconductivity.

**Acknowledgements**: We are grateful to A. Chubukov, S.D. Edkins, M.H. Hamidian, J. E. Hoffman, E.-A. Kim, S. A. Kivelson, M. Lawler, D.-H. Lee and J.-H. She for helpful discussions and communications. J.C.S.D. acknowledges gratefully support from the Moore Foundation's EPiQS Initiative through Grant GBMF4544, and the hospitality and support of the Tyndall National Institute, University College Cork, Cork, Ireland. PJH acknowledges support DOE Grant No. DE-FG02-05ER46236. AKr and BMA acknowledge support from a Lundbeckfond Fellowship (Grant No. A9318). Material synthesis and detailed characterization at Ames National Laboratory was supported by the U.S. Department of Energy, Office of Basic Energy Science, Division of Materials Sciences and Engineering - Ames Laboratory is operated for the U.S. Department of Energy by Iowa State University under Contract No. DE-AC02-07CH11358; Experimental studies were carried out by the Center for Emergent Superconductivity, an Energy Frontier Research Center, headquartered at Brookhaven National Laboratory were funded by the U.S. Department of Energy under DE-2009-BNL-PM015. The data described in the paper are archived by the Davis Research Group at Cornell University.

**Figure 1 Bogoliubov Quasiparticle Interference Model for FeSe**

**(A)** Top view of FeSe crystal structure. Dashed lines represent the 1-Fe unit cell, and the actual unit cell is shown using solid lines. The unit cell of FeSe is distorted in the nematic phase with a$_{Fe}$>b$_{Fe}$. Throughout this paper we define the $x$-axis, $\vec{k}_x$-axis and $\vec{q}_x$-axis to all be parallel to $a_{Fe}$-axis, so that labels of orbitals like $d_{xy}$ or $d_{yz}$ or $\vec{k}$-space locations and states, are equally valid in both nematic domains.

**(B)** In the nematic phase with orthorhombic crystal symmetry, the FeSe Fermi surface consists of a hole-like α-band around Γ=(0,0) and an electron-like ε-band around X=($\pi/a_{Fe}$,0); the color code indicates the regions of Fermi surface dominated by states with primarily $d_{yz}$ (green), $d_{xz}$ (red) and $d_{xy}$ (blue) orbital character. An anticipated third band, the δ-band around Y=(0,$\pi/b_{Fe}$) has not yet been observed by spectroscopic techniques.

**(C)** Constant-energy-contours (CEC) of Bogoliubov quasiparticles for the gapped α-band around Γ=(0,0). The CEC are color-coded to indicate increasing energy. A schematic, ellipsoidal normal state Fermi surface is shown using a grey dashed contour. Predominant scattering interference occurring between the 'tips' of the CEC should produce a triplet of characteristic BQPI wavevectors $\vec{q}_1^\alpha(E), \vec{q}_2^\alpha(E), \vec{q}_3^\alpha(E)$ (black arrows).

**(D)** Same as (C) but for the gapped ε-band around X=($\pi/a_{Fe}$,0).

**(E)** The expected energy dependence of the α-band wavevector triplet $\vec{q}_1^\alpha(E), \vec{q}_2^\alpha(E), \vec{q}_3^\alpha(E)$ in (C); these are color-coded to indicate increasing energy. The black diamond symbolizes the starting point of $\vec{q}_3^\alpha$ where $\Delta_\alpha = max$.

**(F)** The expected energy dependence of the ε-band wavevector triplet $\vec{q}_1^\varepsilon(E), \vec{q}_2^\varepsilon(E), \vec{q}_3^\varepsilon(E)$ color-coded by energy. The black diamond symbolizes the end point of $\vec{q}_2^\varepsilon$ where $\Delta_\varepsilon = min$.

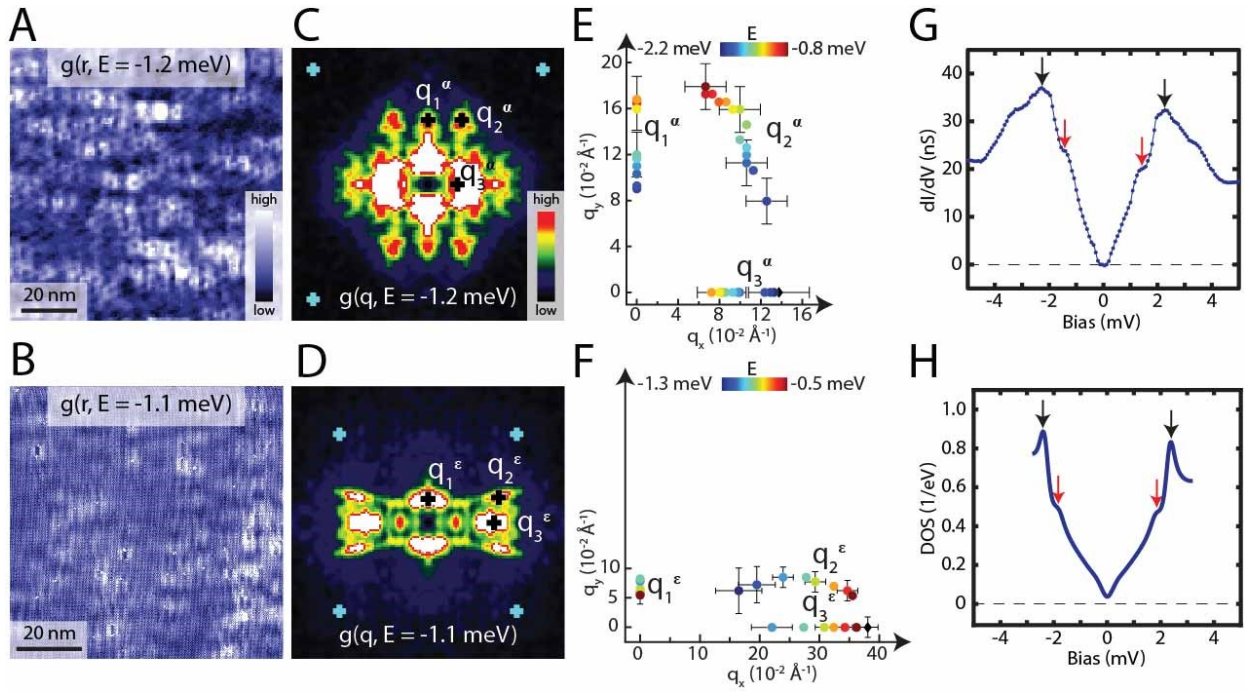

**Figure 2      Visualizing Bogoliubov Quasiparticle Interference in FeSe**

To achieve sufficient $\vec{q}$-space resolution we use an approximately 90 nm X 90 nm field-of-view and image $g(\vec{r},E) \equiv dI/dV(\vec{r}, E = eV)$ with bias modulation of 100 µV at T=280mK (section IV of SM).

(A) Typical measured $g(\vec{r},E)$ using a low resolution STM-tip which predominantly is sensitive to α-band effects.

(B) Typical $g(\vec{r},E)$ measured with very high spatial resolution tips, which emphasize very short wavelength BQPI, and are predominantly sensitive to ε-band effects.

(C) Measured $|g(\vec{q},E)|$ derived from (A) where the BQPI wavevector triplet $\vec{q}_i^\alpha(E)$ is identified by black crosses at the points of maximum amplitude. The blue crosses indicate the $(\pm 2\pi/8a_{Fe}, \pm 2\pi/8b_{Fe})$ points.

(D) Measured $|g(\vec{q},E)|$ derived from (B) where the BQPI triplet $\vec{q}_i^\varepsilon(E)$ is identified by black crosses at the points of maximum amplitude. The blue crosses indicate the $(\pm 2\pi/8a_{Fe}, \pm 2\pi/8b_{Fe})$ points.

(E) Measured evolution of $\vec{q}_1^\alpha(E), \vec{q}_2^\alpha(E), \vec{q}_3^\alpha(E)$. The $|g(\vec{q},E)|$ data is shown in movie S1. The black diamond is the first $\vec{q}_3^\alpha(E = -2.3\ meV)$ data point.

(F) Measured evolution of $\vec{q}_1^\varepsilon(E), \vec{q}_2^\varepsilon(E), \vec{q}_3^\varepsilon(E)$. The $|g(\vec{q},E)|$ data is shown in movie S2. The black diamond corresponds to the last $\vec{q}_2^\varepsilon(E = -0.3\ meV)$ data point.

(G) Measured $N(E)$; black arrows indicate the maximum energy gap on any band, which we determine from BQPI to be on the α-band (Fig. 3). Red arrows indicate a smaller energy gap on a second band which from BQPI is assigned to the ε-band (Fig. 3).

(H) Calculated $N(E)$ from the band-structure and gap structure model (supplementary text section II and VIII).

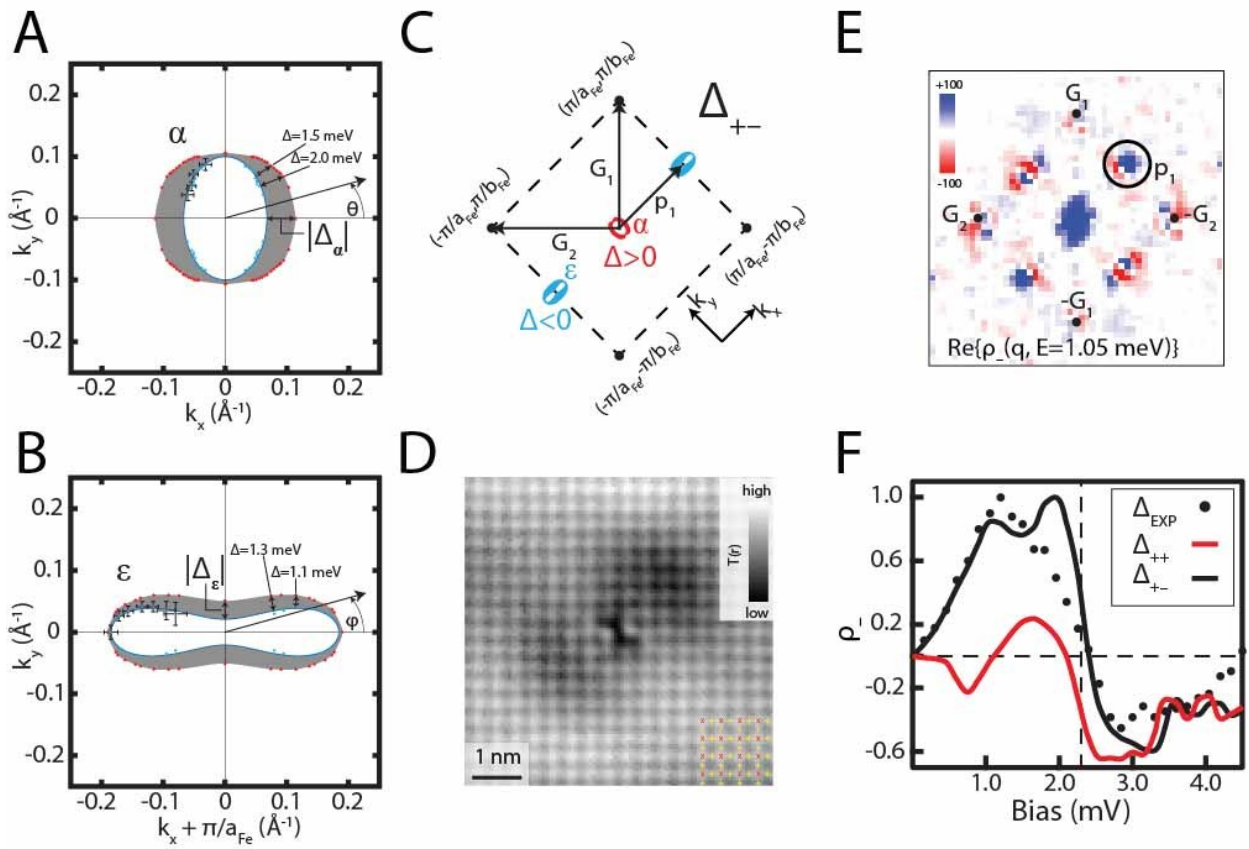

**Figure 3  BQPI Determination of Fermi Surfaces and Energy Gaps**

The BQPI data analysis steps yielding the results in Fig. 3 is explained in full detail in section V of SM.

(A) The Fermi surface of α-band is measured using the BQPI triplet $\vec{q}_1^\alpha(E), \vec{q}_2^\alpha(E), \vec{q}_3^\alpha(E)$ and shown as black dots with error bars. Energy-gap magnitude for the α-band also measured using energy dependence of the BQPI triplet $\vec{q}_1^\alpha(E), \vec{q}_2^\alpha(E), \vec{q}_3^\alpha(E)$ plus the values of maximum and minimum energy gap from N(E) in 2G.

(B) Fermi surface of ε-band measured using the BQPI triplet $\vec{q}_1^\varepsilon(E), \vec{q}_2^\varepsilon(E), \vec{q}_3^\varepsilon(E)$ and shown as black dots with error bars. Energy-gap magnitude for the ε-band measured using energy dependence of the BQPI triplet $\vec{q}_1^\varepsilon(E), \vec{q}_2^\varepsilon(E), \vec{q}_3^\varepsilon(E)$ and maximum/minimum energy gap from N(E) in 2G.

(C) $\vec{k}$-space schematics of FeSe interband scattering wavevector $\vec{p}_1$ between α- and ε-bands which connects gaps of opposite sign in the $\Delta_{+-}$ scenario.

(D) Measured $T(\vec{r})$ topograph centered on a typical individual impurity site in a ~ 6.5x6.5 nm² field of view. The surface (upper) Se sites are shown using red x symbol, and the Fe sites using yellow +.

(E) Typical measured $\rho_-(\vec{q}, E) = Re\{g(\vec{q}, +E)\} - Re\{g(\vec{q}, -E)\}$ from BQPI $g(\vec{r}, E)$ at E=1.05meV in the energy range within both $\Delta_\alpha$ and $\Delta_\varepsilon$. Complete $\rho_-(\vec{q}, E)$ is shown in movie S3.

(F) Predicted $\rho_-(E)$ for ± pairing symmetry using the band/gap structure of FeSe (sections II, VI and VIII of SM and Fig 3), shown as solid black curve. The measured $\rho_-(E)$ for FeSe (black dots) is calculated by integration over the $\vec{q}$-space region identified by the black circle in 4E. Predicted $\rho_-(E)$ for no gap sign change in FeSe shown as solid red curve. The vertical dashed black line marks the energy of the maximum superconducting gap.

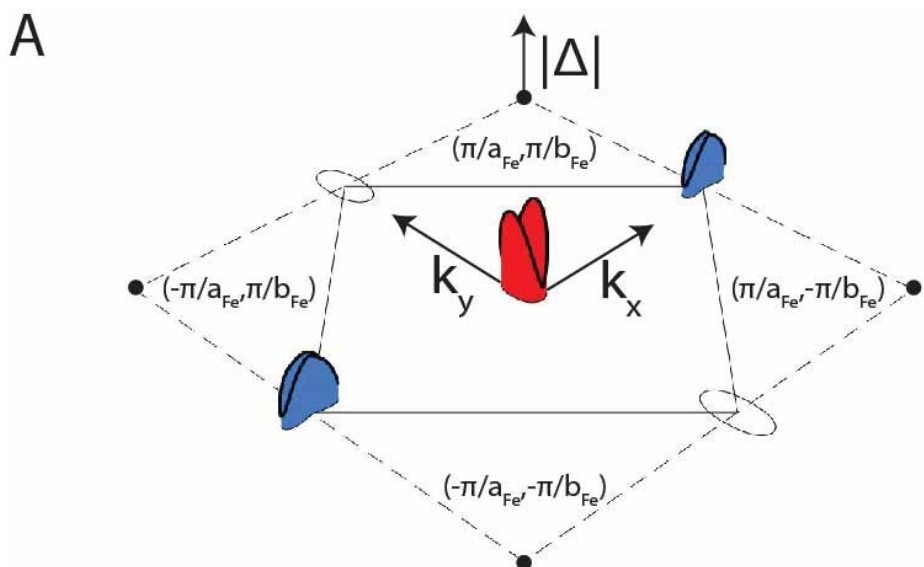
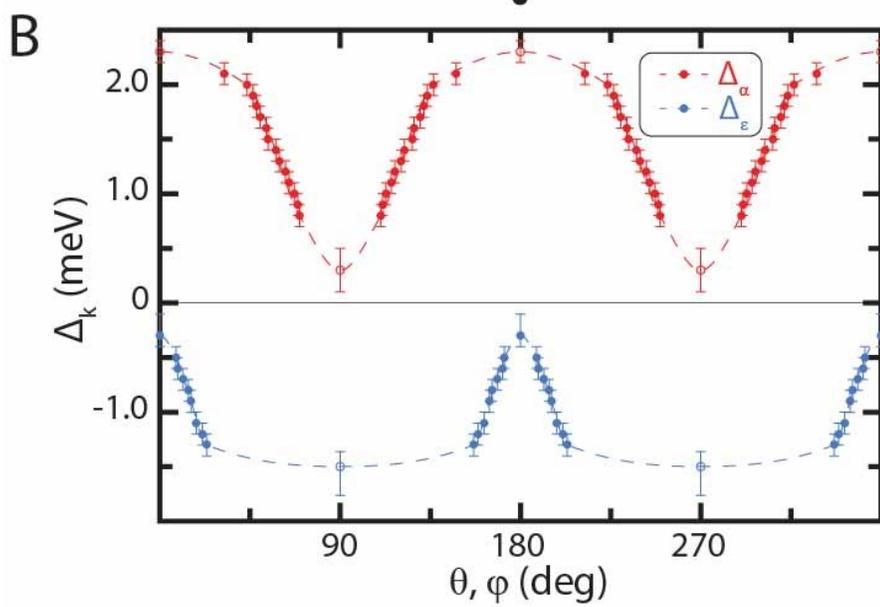
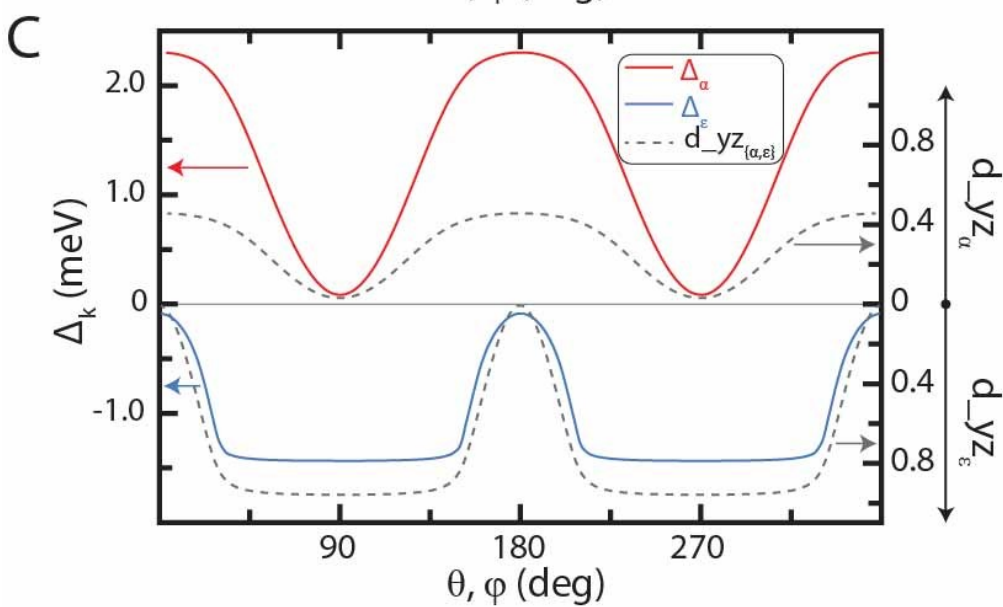

**Figure 4   Orbitally Selective Cooper Pairing in FeSe**

A. Measured $\vec{k}$-space structure of anisotropic energy gaps of FeSe (Fig. 3). The red and blue colors indicate the different signs of the two gap functions.

B. Measured angular dependence of FeSe superconducting energy-gaps $\Delta_\alpha(\vec{k})$ about $\Gamma=(0,0)$ and the equivalent for $\Delta_\varepsilon(\vec{k})$ about $X=(\pi/a_{Fe},0)$ from Fig. 3A,B.

C. Predicted angular dependence of $\Delta_\alpha(\vec{k})$ and $\Delta_\varepsilon(\vec{k})$ for an interband pairing interaction that is peaked at $\vec{q}=(\pi/a_{Fe},0)$ and for which pairing is orbital selective, occurring predominantly for electrons with $d_{yz}$ orbital character (section VIII of SM). The dashed grey curves show the $d_{yz}$ orbital character of states at the $\alpha$-band and $\varepsilon$-band Fermi surfaces.

# Supplementary Materials for
# Discovery of Orbital-Selective Cooper Pairing in FeSe

P.O. Sprau, A. Kostin, A. Kreisel, A. E. Böhmer, V. Taufour, P.C. Canfield,
S. Mukherjee, P.J. Hirschfeld, B.M. Andersen and J.C. Séamus Davis

correspondence to: jcseamusdavis@gmail.com

## Materials & Methods

The FeSe single crystals were prepared using KCl/AlCl$_3$ chemical-vapour transport (*1*) and were thoroughly characterized using resistivity, magnetization and x-ray diffraction measurements (*1-4*). They show a structural transition at $T_s$ = 87–89 K and a superconducting transition at $T_c$ = 8.7–8.8 K.

For study of the samples in the Spectroscopic Imaging-Scanning Tunneling Microscope (SI-STM), each single crystal is glued flat onto the end of a cylindrical brass sample holder using silver epoxy H20E from Epotek. This provides excellent heat and electrical conductivity at low temperatures and yields a clean flat unstressed cleave of the crystal with high reliability. All samples are inserted slowly from a room temperature load-lock into the cryogenic environment and then cleaved in situ in cryogenic ultra-high vacuum at $T < 20\ K$.

## Supplementary Text

### I. CRYSTAL STRUCTURE OF FeSe

The high temperature structure of FeSe belongs to the tetragonal *P4/nmm* space group with the corresponding lattice parameters *a*= 3.77 Å and *c*= 5.52 Å (*5,6,7*). Note that here *a* is the distance between nearest neighbor selenium atoms lying in the same plane. The Fe-Fe nearest neighbor distance is then $a/\sqrt{2}$. On cooling, FeSe undergoes a transition to *Cmma* orthorhombic structure with the lattice parameters *a*= 5.31 Å, *b*= 5.33 Å and *c*= 5.48 Å (*5,7*). Here *a* and *b* correspond to the two next nearest neighbor distances between selenium atoms lying in the same plane (Fig. S1). In the orthorhombic phase, the two inequivalent Fe-Fe distances are related to the lattice parameters via $a_{Fe} = b/2$ and $b_{Fe} = a/2$. These distances are the most convenient parameters for the discussion of the electronic structure of FeSe, and hence we adopt them throughout this paper. The convention $a_{Fe} > b_{Fe}$ is chosen to match earlier work (8,9) and for clarity of communication of our key results, even though for the orthorhombic *Cmma* lattice parameters $a < b < c$ is usually enforced.

### II. TIGHT-BINDING MODEL FOR FeSe

For the present investigation, we use a band structure model, introduced in (10,11), that includes a parameterization of site-centered ($\Delta_s$) and bond-centered ($\Delta_b$) orbital order as well as spin-orbit coupling (12). Effects of interactions and correlations of the quasiparticles (in the normal state, but including the nematic order) are modeled by allowing the hoppings to be modified to match the spectral positions of the quasiparticle peaks observed in ARPES (9,13,14) and BQPI (this work). Note that this approach takes into account the real part of the self-energy corrections by fitting to the experimentally observed spectral positions $E_i(\mathbf{k})$. Specifically, the band structure is determined by the normal state Hamiltonian $H_N = H_0 + H_{OO} + H_{SOC}$, where $H_0$ (in real space notation) is given by

$$H_0 = \sum_{r,r',a,b} t^{ab}_{r-r'} c^{\dagger}_{a,r} c_{b,r'} \tag{S1}$$

where $a, b$ are orbital labels and $\mathbf{r}, \mathbf{r}'$ are lattice sites. For the orbital order term, we use the momentum space representation,

$$H_{OO} = \Delta_b(T) \sum_k (\cos(k_x) - \cos(k_y))(n_{xz}(\mathbf{k}) + n_{yz}(\mathbf{k})) + \Delta_s(T) \sum_k (n_{xz}(\mathbf{k}) - n_{yz}(\mathbf{k})) \tag{S2}$$

Finally, the spin orbit coupling is given by

$$H_{SOC} = \lambda \mathbf{L} \cdot \mathbf{S} \tag{S3}$$

Keeping in mind the correspondence between a 5-band and a 10-band models (*11*) which is exact for $k_z = 0$ and $k_z = \pi$ in the absence of spin-orbit coupling (SOC), we use a 5-band model to represent the band structure away from the band crossings that have splittings induced by SOC. However, the splitting at $\Gamma$ needs to be taken into account (*15,16*). The hoppings for the tight-binding model are therefore given in the 5-band representation in the separate attached file.

The resulting band structure is shown in Fig. S2. It consists of a hole-like $\alpha$-band with ellipsoidally shaped Fermi surface surrounding $\Gamma=(0,0)$ whose major axis is aligned to the orthorhombic $b_{Fe}$-axis; an electron-like "bow-tie"-shaped $\varepsilon$-band surrounding X=$(\pi/a_{Fe},0)$ whose major axis is aligned to the orthorhombic $a_{Fe}$-axis; an electron-like $\delta$-band surrounding Y= $(0, \pi/b_{Fe})$. We define the *x*-axis (*y*-axis) to always be parallel to the orthorhombic $a_{Fe}$-axis ($b_{Fe}$-axis).

The values of the orbital order terms, as determined from experimental measurements of normal state QPI to be discussed elsewhere, are $\Delta_s = 9.6$ meV, $\Delta_b = -8.9$ meV, and the SOC constant is fixed to $\lambda = 20$ meV. These values also agree with the observed splitting above the nematic ordering temperature (*16*) assuming that the SOC is unaffected by temperature.

Because they are key to interpreting the data, we show in Fig. S3 the $k_z=0$ Fermi surface (FS) of our band structure and the contributions of the $d_{xz}$, $d_{yz}$, and $d_{xy}$ orbitals on the FS. Greater line thickness corresponds to larger orbital contribution throughout.

In order to visualize the influence of the orbital order terms on the Fermi surface topology, we display the 10-band spectral functions at $\omega = 0$ both above and below the structural transition $T_S$ which correspond to $\Delta_s = 0$ meV, $\Delta_b = 0$ meV and $\Delta_s = 9.6$ meV, $\Delta_b = -8.9$ meV, respectively (Fig. S4). Above the structural transition, without orbital order, the Fermi surface is $C_4$-symmetric as reported by ARPES Ref. 14.

Overall, our band structure parametrization not only agrees in its spectral positions $E_i(k)$ with those of experimental observations, but is also consistent with deductions of the orbital content of the Fermi surface (*9*). Moreover, at low energies it does not show any unexpected behavior as compared to investigations that include correlations (*17*). Its general correspondence with experiment is discussed in detail in the next section.

### III. CONSISTENCY OF BQPI, QO AND ARPES FOR FeSe FERMI SURFACES

This model band structure (Section II) is in close agreement with key experimental observations including:

#### A. BQPI and ARPES at $k_z = \pi$

Figure S5 shows directly the agreement of our model band structure to the position of the Fermi surface that we measure using BQPI (red dots), as presented in the main text Fig. 3. The black line is the calculated / tight-binding / model Fermi surface at $k_z = 0$ from our band-structure (Section II); it agrees with the measured positions within the experimental error bars. At the same time, the dispersion of our band-structure along $k_z$ is significant and agrees well with the findings from a recent ARPES investigation (*13*). In Fig S5, we also reproduce the spectral function measured by ARPES (*14*) at $k_z = \pi$, and it compares very well to the model Fermi surface at $k_z = \pi$ (blue line). The superposition of copies of the basic Fermi surface spectral-function features but rotated by π/2, is due to the existence of both orthorhombic domains in the ARPES study; obviously our band structure model does not reproduce these effects. Note that the same experimental work also revealed the Fermi surface at $k_z = 0$ which gives very similar spectral positions as the ones obtained from BQPI. Therefore, we concentrate on the use of the model system for $k_z = 0$ whenever carrying out simulations for comparison with the data.

#### B. Quantum oscillations

For the calculation of the extremal Fermi surface areas we use the 10-band analog of our band structure (*11*) that additionally introduces small hybridizations on the electron pockets yielding an orbit of the "inner" electron pocket and the "outer electron" pocket. The extremal areas obtained within this method would give rise to quantum oscillation frequencies of: 66T (inner electron pocket), 199 T (hole pocket at $k_z = 0$), 579T (outer electron pocket), and 651 T (hole pocket at $k_z = \pi$); results were rounded to last digit. The measured extremal frequencies for magnetic field angle $\theta = 0$ are reported as values in the range of 60 T – 114 T, 200 T – 207 T, 530 T – 580 T, and 660 T – 680 T, see table I in Ref. (*13*), and references therein. Thus, there is good comprehensive agreement within the experimental uncertainties between the band structure described in Fig. S2 and the quantum oscillation data.

## IV. BQPI DATA FOR FeSe
### A. BQPI Data Acquisition

Differential tunneling conductance $dI/dV(\vec{r}, eV) \equiv g(\vec{r}, E)$ was measured at T=280mK, and as a function of both location $\vec{r}$ and electron energy $E=eV$ where V is the tip-sample bias voltage. We use fields of view in the range of 60nm×60nm to 90nm×90nm square and raster between 128x128 to 400x400 pixel square to get high signal-to-noise ratio and sufficient $q$-space resolution, and a typical bias modulation of δV=100 μeV. The same measurements were carried out on both nematic domains of multiple crystals and all the results presented herein are supported by this data set.

Three types of tips were repeatably observed during these studies. Figure S6 presents conductance maps $g(r, E)$ at -1.1 meV measured with the three different tip types (for clarity shown in smaller fields of view in q-space than the original data) along with the amplitude of their corresponding Fourier transforms $|g(q, E)|$. The tips differ in their 'atomic' sharpness, with the 'sharper' tips being created when the atomic configuration at the end of the tip changed while scanning across the surface at low tunneling junction resistance. Figs. S6A and S6B represent the expected case where the STM tip is sensitive to the BQPI signal from both the α- and ε-pocket simultaneously. It is possible in principle to extract the FS and energy gaps $\Delta_\alpha(\vec{k})$ and $\Delta_\varepsilon(\vec{k})$ from these data alone, and our measurements of them are in good agreement with the results presented in the main text. However, we found that it is also possible to simplify the situation and to measure the properties of the two bands individually. Figs. S6C and S6D contain the results for a tip that is sensitive primarily to the α-pocket. This is achieved by using a lower spatial resolution tip which is obviously far more sensitive to the long wavelength BQPI modulations that occur in intra-band scattering in the center of the BZ. Alternatively, Figures S6E and S6F show the results form a tip that is predominantly sensitive to the scattering interference from the ε-pocket. Momentum dependence of tunneling tips is discussed in more detail in Ref. (*18*). In the following we will call these tips 'tip αε', 'tip α', and 'tip ε'. All panels in Fig. S6 and S7 have been labeled depending on which tip was used during the measurement.

As can be seen from the atomic contrast in Figs. S6A, S6E the tips sensitive to the the ε-pocket BQPI possess excellent real-space resolution. In order to compare real-space resolution and q-space sensitivity to the tip that is primarily sensitive to the α-pocket BQPI we show in Fig. S7A, S6B constant current topographs recorded for the same setup current and setup bias but with two different tips: 'tip α' and 'tip αε'. The topographs and the corresponding amplitudes of their Fourier transforms in Fig. S7C, S7D exhibit the same behavior as seen in Fig. S6. Superior real space resolution goes hand in hand with sensitivity to high-q phenomena as expected. The 'sharper' tip is sensitive to even the signal from 'Umklapp'-scattering processes around the Se-Bragg peak which has been marked with a white circle. Additionally, the 'sharper' tip is sensitive to interband scattering between the electron and the hole pocket at $(\frac{\pi}{a_{Fe}}, 0)$ which is absent for the lower resolution α-tip, see Fig. S7C. We find identical high-q properties for the atomically 'sharp' tip which predominantly detects the ε-pocket for low q-values, 'tip ε'.

We assign the BQPI triplet of wavevectors $\vec{q}_i^{\alpha,\varepsilon}(E)$ to the α- and ε-pocket based on the energy evolution of their intensity-maxima, which can be compared to JDOS (Joint Density of States) simulations of the expected BQPI using our tight-binding model in Fig. 1. The JDOS simulations will be discussed at the end of this section in further detail, and the combined $g(\vec{r}, E)$, $|g(\vec{q}, E)|$ data plus relevant JDOS simulations are added as supplementary movies S1 and S2.

## B. BQPI Data Processing Steps

In Fig. S8 we present the sequence of steps used to optimize the signal-to-noise ratio of the BQPI data. The three steps are symmetrization, averaging, and Gaussian core subtraction in Fourier space. We would like to emphasize that no unfolding of BQPI data with respect to the 1-Fe- and 2-Fe-unit cell picture takes place within these steps, and that the symmetrization does not enforce $C_2$ symmetry on the data.

For the symmetrization we take advantage of the mirror symmetry axes present in the underlying k-space structure, see also Fig. S4. As BQPI consists of scattering between parts of constant energy contours in k-space these mirror symmetry axes carry over into q-space. The raw amplitude Fourier transform $|g(\vec{q}, E)|$ is symmetrized via reflection about the mirror symmetry axes displayed in Fig S8A.

In order to further increase the signal-to-noise ratio we use a three-by-three pixel averaging filter on the symmetrized amplitude Fourier transform $|g(\vec{q}, E)|$ in Fig. S8B. The result is shown in Fig. S8C. The last step is a Gaussian core subtraction in Fourier space which corresponds to a long wavelength filter in real space. As can be seen in Fig. S8D this subtracts intensity for very small q-vectors.

## C. JOINT DENSITY OF STATES BQPI SIMULATIONS

In order to compare the observed BQPI to the proposed tight-binding and pairing model we simulate $JDOS(\boldsymbol{q}, \omega) = \int A(\boldsymbol{k} + \boldsymbol{q}, \omega) A(\boldsymbol{k}, \omega) d\boldsymbol{k}$ with the spectral function $A(\boldsymbol{k}, \omega)$ given by $A(\boldsymbol{k}, \omega) = -\frac{1}{\pi} Im\{\sum_a G_{aa}(\boldsymbol{k}, \omega)\}$, where the sum runs over the orbitals $a$. Here $G_{aa}(\boldsymbol{k}, \omega) = Z_a G_{aa}^0(\boldsymbol{k}, \omega)$ is the dressed Green's function, and the Z-factors for the orbitals are the same as used for the calculation of the pairing interaction (see section VIII for more details), and given as $Z_a \in \{0.2715^2, 0.9717^2, 0.4048^2, 0.9236^2, 0.5916^2\}$. Furthermore, we separate the JDOS into partial JDOS simulations by restricting the integration area to ¼ of the Brillouin zone containing the corresponding pocket. Here to mimic the sensitivity of our tunneling tips to BQPI from different bands, we compute the partial JDOS for the α-pocket, for the ε-pocket, and the sum of both. This is a valid approach as the two pockets are well-separated in k-space. The results in Fig. S9 are in excellent agreement with the 'banana tips' model, as it clearly visualizes the three (independent) dominant scattering vectors $\{\boldsymbol{q}_1, \boldsymbol{q}_2, \boldsymbol{q}_3\}$ for each pocket, which are transformed into a set of eight through symmetry operations. Overall we find very good agreement between experiment and JDOS simulation, and deviations between the two can for example be ascribed to the static structure of the scatterers themselves which the JDOS simulation cannot take into account. The relevant energies for the JDOS simulations are part of the supplementary movies S1 and S2. In addition to the movies, we present typical measured BQPI data at four energies, and comparison to partial JDOS, for both the α - and ε-pocket in Fig. S10 and Fig. S11.

## V. EXPERIMENTAL DETERMINATION OF FS AND ENERGY GAPS

Consider a standard Bogoliubov spectrum of a superconductor with an anisotropic gap.

$$\epsilon_k = \pm\sqrt{E_k^2 + \Delta_k^2} \tag{S4}$$

Without loss of generality, we can then define a constant-energy-contour (CEC) in k-space at a specified energy $\epsilon' > 0$ (because of the particle-hole symmetry, all observations also apply to $-\epsilon' < 0$) by the following equation:

$$\epsilon' = \sqrt{E_k^2 + \Delta_k^2} \qquad (S5)$$

Let's further impose that the contour is within the superconducting gap meaning $\epsilon' < \Delta_{max}$. As long we are in the region of k space where $\epsilon' > \Delta_k$, we expect two types of k-space solutions, one with $E_k < 0$ and one with $E_k > 0$. These two types of solutions will connect at specific k points where $\epsilon' = \Delta_k$, and hence $E_k = 0$, to create closed contours reminiscent of bananas. (See Fig. S12A below.) Since $E_k = 0$, these points ('banana-tips') lie on the normal state FS of the corresponding band by definition.

Within the JDOS picture, the modulations in the density of states (and hence dI/dV) due to impurity scattering will be dominated by the q-vectors connecting k-space regions with high spectral weight. For a superconductor with an anisotropic gap, such regions are exactly the tips of CEC discussed above (Fig. S12B). For that reason, tracking the evolution of the q-vectors associated with the tips of the CEC allows one to extract both the Fermi surface and the k-space structure of the gap.

Thus for FeSe, using the measured BQPI wavevector sets $\vec{q}_i^{\alpha,\varepsilon}(E)$ shown in Fig. 3, we determine the two FS by using Eqn. 1 to find the $(k_x,k_y)_{\alpha,\varepsilon}$ locations of all the 'banana-tips' for both the α- and ε-bands. These FS are shown in Fig. 3 of the main text. Next, we use the same wavevector sets $\vec{q}_i^{\alpha,\varepsilon}(E)$ in conjunction with the two FS, to plot the energy $E = \Delta$ associated with the observation of BQPI for each FS wavevector $(k_x,k_y)_{\alpha,\varepsilon}$. The resulting functions are $\Delta_\alpha(\vec{k})$ and $\Delta_\varepsilon(\vec{k})$.

In the 'banana-tips' model, the dispersion of the BQPI wavevector sets $|\vec{q}_i^{\alpha,\varepsilon}(E)|$ is determined by the topology of the Fermi surface and the k-space structure of the gap as discussed above. Because all wavevectors of a set are interdependent due to the geometric restrictions generated by the shape of the Fermi surface, it is possible to check that the extracted BQPI wavevector sets $\vec{q}_i^{\alpha,\varepsilon}(E)$ are internally consistent with one another. Figure S13 demonstrates this consistency. There is very good agreement between the dispersion of the extracted BQPI wavevector sets $|\vec{q}_i^{\alpha,\varepsilon}(E)|$ and the expected dispersion based on the 'banana-tips' model.

## VI. HAEM PREDICTIONS FOR SIGN CHANGE BETWEEN GAPS

Following Ref. (*19*), we perform a calculation of the BQPI response within the T-matrix approach. For this purpose, the Nambu Hamiltonian $H = \begin{pmatrix} H_N & \Delta \\ \Delta^T & -H_N^T \end{pmatrix}$ as a matrix in orbital space is set up where the Hamiltonian is given in section II. The superconducting gap is taken from a self-consistent calculation, (*20,21*) using the same band and pairing as outlined in the main text Figs 3,4 and section VIII. Next, a weak (attractive) nonmagnetic impurity is modeled as a potential scatterer on a single Fe position, motivated by the Fe centered defects seen in the present Spectroscopic Imaging – Scanning Tunneling Microscopy (SI-STM) experiment and also by other groups (*22,23*). These defects are also observed in the monolayer FeSe (*24*). Setting $H_{imp} = V_0 \delta_{ab}$ as a constant on-site scatterer at position $r^*$ in orbital space, we use the additional impurity term in the Hamiltonian $H_{imp} = V_0 \sum_a c_{a,r^*}^\dagger c_{a,r^*}$ which describes the scattering from an impurity centered at a Fe lattice position in the approximation of a short-range potential. Then we solve the T-matrix using the local Green's function $G_0(\omega) = \sum_k G_k^0$ where $G_k^0 = (\omega + i0^+ - H_k)^{-1}$. Noting that the impurity potential is constant in momentum space, we obtain the Green's function in the presence of scatterer as $G_{k,k'}(\omega) = G_{k-k'}^0(\omega) + G_k^0(\omega) T(\omega) G_{k'}^0(\omega)$. The T-matrix is obtained from the equation $T(\omega) = [1 - V_{imp} G_0(\omega)]^{-1} V_{imp}$ such that the change in the local density of states is given by $\delta N(\boldsymbol{q}, \omega) = \frac{1}{\pi} Tr\{Im \sum_k G_k^0(\omega) T(\omega) G_{k+q}^0(\omega)\}$.

Theoretically, the quasiparticle scattering between states with sign-changing order parameter yields a characteristic resonant energy dependence in $\rho_-(\omega)$, the anti-symmetrized QPI response, integrated over a finite momentum space corresponding to relevant inter-band scattering processes as discussed in Ref. (*19*). To pick out only the inter-band scattering contributions in FeSe which are sign-changing in the $\Delta_\pm$ scenario, we integrate over an area in momentum space centered at $(\pi, 0)$ to obtain $\rho(\omega) = \sum_q' \delta N(\boldsymbol{q}, \omega)$ and construct $\rho_-(\omega) = \rho(\omega) - \rho(-\omega)$. These calculations are done for two different gap structures that yield the same density of states because they differ only by a relative sign between the order parameters on the electron band and the hole band. The results for the calculations are presented in Fig. 3F. The quantity $\rho_-(\omega)$ changes sign at an energy within the superconducting gap for the sign-preserving order parameter $\Delta_{++}$. For the sign-changing order parameter $\Delta_{\pm}$, there is however no change of sign in $\rho_-(\omega)$ up to roughly the maximum superconducting gap, see Figs. 3F, S15E and S15F.

## VII. SINGLE IMPURITY SITE MEASUREMENTS IN FeSe

### A. Extracting $\rho_-(\omega)$ from differential tunneling conductance measurements

In order to extract $\rho_-(\omega)$ from the experiment we take advantage of the fact that the differential tunneling conductance $g(\vec{r}, \omega)$ is proportional to the local density of states of the sample $\rho(\vec{r}, \omega)$. Next we construct in Fourier space the real part of the anti-symmetrized differential tunneling conductance $\rho_-(\vec{q}, \omega) = Re\{g(\vec{q}, +\omega)\} - Re\{g(\vec{q}, -\omega)\}$ and integrate in a circular region around $(\pi, 0)$-scattering which connects the electron and hole pocket at $X$ and $\Gamma$, respectively: $\rho_-(\omega) = \sum_{(q_x - p_{1,x})^2 + (q_y - p_{1,y})^2 \leq \delta q^2} \rho_-(\vec{q}, \omega)$. Here $\vec{p}_1 = (p_{1,x}, p_{1,y})$ corresponds to the position of $\left(\frac{\pi}{a_{Fe}}, 0\right)$-scattering in **q**-space, and the radius used for integration $\delta q$ is chosen so that we capture only intensity related to scattering between the electron and hole pocket at $X$ and $\Gamma$. $\delta q$ is thus determined by the size of the two pockets.

However, before one computes $\rho_-(\omega)$, any shift of the scatterer away from the origin of the Fourier transform (FT) needs to be corrected as exactly as possible. The correction is necessary as any shift of the scatterer, i.e. the impurity, in real space away from the origin of the FT creates an additional phase term in **q**-space according to the shift theorem of FTs: $FT\{f(\vec{r} - \vec{r}_0)\} = e^{-i\vec{q} \cdot \vec{r}_0} FT\{f(\vec{r})\}$. Fortunately, the shift theorem allows one to correct the data. The experimental challenge lies therefore in determining the spatial position of the scatterer with high precision.

As mentioned in section VI, the dominant type of defect in FeSe is centered on an Fe-atom. We know the relative position of Se- and Fe-atoms from the crystal structure. FeSe cleaves between two layers of selenium so that it is a reasonable assumption that SI-STM images the Se-atoms. Before we determine the spatial shift of the impurity we correct the data for distortions due to non-orthogonality in the x/y axes of the piezoelectric scanner tube using the Lawler-Fujita algorithm (*25*). After that we shift the data to the pixel corresponding to the origin of the FT.

The measured $\rho_-(\omega)$ is shown in Fig. S14. To increase the signal-to-noise ratio we averaged over adjacent energy values of $\rho_-(\omega)$. Note that the small Fermi surface pockets clearly separate scattering between states from and to different pockets, e.g. there are no intraband scattering events at the relative energy scale that are picked up by the integral. Resolving the structure of the tiny Fermi surface pockets is an experimental challenge because large FOV are required in STM, but at the same time it allows one to clearly separate intraband contributions. In order to test how robust the result is under change of the integration area, we used three circular areas as depicted in Fig. S14D. The radius for the smallest, blue circle was three quarter the size of the medium black circle, and the radius of the biggest, magenta circle was five quarters the size of the medium black circle. As can be seen in Fig. S14F, the results are very consistent, and the biggest change occurs for energies outside the superconducting gap. For clarity and easier comparability $\rho_-(\omega)$ has been normalized to its maximum value for all three cases.

## B. Repeatability of HAEM Procedure

In order to test the generality of the result for $\rho_-(\omega)$ in section VII A we repeat the analysis for two more impurities. The second impurity was in the same sample studied with the same tip, and its relative position to the first impurity can be seen in Fig. S15A-C. The third impurity was in a second sample, and was examined with a physically different tip. Its position is marked in the topograph in Fig. S15D.

For both additional impurities the results agree best with a sign-changing superconducting order parameter, see Figs. S15E and S15F. In order to make the visual comparison of experiment and theory easier, $\rho_-(\omega)$ has been normalized to its maximum value for all three impurities and the theoretically predicted case of $\Delta_{+-}$. Theoretically predicted $\rho_-(\omega)$ for the case of $\Delta_{++}$ has been normalized so that the relative magnitude to the theoretically predicted case of $\Delta_{+-}$ is conserved. There is no signature of a sign change in $\rho_-(\omega)$ within $\omega \leq \Delta$ which is a clear indication for the $\Delta_{+-}$ scenario and rules out the sign-preserving order parameter within our modeling.

Importantly, the $g(\vec{r},\omega)$ and $\rho_-(\omega)$ at the impurity studied in a second sample with a physically different tip, agrees remarkably well with the first, isolated impurity. From that we conclude that the result supporting sign-changing superconducting order is both very robust and intrinsic to the material studied.

### C. Defect induced in-gap states and disappearance of $\rho_-(\vec{q},\omega)$-signal above T$_C$

We additionally analyzed the single defect by taking dI/dV line-cuts through it along the x- and y-direction. Fig. S16 presents dI/dV line-cuts through the single defect. The differential conductance inside the superconducting gap exhibits an increase at the defect location. The effect is small, but clearly visible.

In order to confirm that the $\rho_-(\vec{q},\omega)$-signal originates from scattering inside the superconducting state, and is not for example due to effects of the band structure or the defect itself we analyze a defect at T = 4.2 K < T$_C$ and T = 10.0 K > T$_C$. Setup voltage (V$_s$ = -50 mV), tunneling current (I = 500 pA), and modulation voltage for the lock-in measurement (V$_M$ = 1 mV) were identical in both measurements. As shown in Fig. S17A-D the signal related to scattering between the α- and ε-pocket vanishes above T$_C$. At the same time no sign of the superconducting gap remains visible in spectroscopy above T$_C$, see Fig. S17E. From that we conclude that the majority of the $\rho_-(\vec{q},\omega)$-signal stems from the properties of the superconducting ordered state in FeSe.

## VIII. MODELING ORBITAL-SELECTIVE COOPER PAIRING IN FeSe

Within a theoretical model for quasiparticles in a correlated material, one can parameterize the full Green's function using a quasiparticle weight $Z_k$ and a modified dispersion $\tilde{E}_k$ as $G(k,\omega) = Z_k/(\omega + i0^+ - \tilde{E}_k)$. Considering our approach to determine the quasiparticle energies by a fit to experimental results from ARPES, QO and SI-STM data for the present material, the corrections to $\tilde{E}_k$ are already taken into account by the shifts and additional terms as outlined in section I.

A calculation of the gap structure within a model that only takes into account the modified quasiparticle energies fails: From standard spin fluctuation theory one obtains a rather small gap on the α-pocket (even slightly smaller than in earlier 3D calculations (*26*)) with limited anisotropy and a large gap on the epsilon pocket, but with opposite anisotropy as shown in Fig. S18A. Furthermore, the spin-fluctuation spectrum is dominated by $(\pi,\pi)$ fluctuations rather than $(\pi,0)$ fluctuations as observed in INS (*27,28*). Neither imposing strong $(\pi,0)$ fluctuations 'by hand', nor using the slightly modified band structure of Ref. (*26*), which actually accounts well for the measured neutron measurements, resolves the discrepancy. We therefore postulate that the missing ingredient is strong orbital selectivity in the quasiparticle weights.

From many-body methods (*17,29*), it is known that the $d_{xy}$ orbital is strongly renormalized, i.e. the quasiparticle weight $Z_k$ is suppressed. In the nematic state one additionally expects that the quasiparticle weight of the $d_{xz}$ orbital is different from that of the $d_{yz}$ orbital. Using this information, one can dress the Green's function as derived from our tight-binding model with a simple multiplication of an orbital-dependent quasiparticle weight:

$$G_{ab}(\mathbf{k},\omega) = Z_{ab} G^0_{ab}(\mathbf{k},\omega) \tag{S6}$$

where $Z_{ab} = \sqrt{Z_a}\sqrt{Z_b}$ is given by the geometric mean of the quasiparticle weights of the connected orbitals in the spirit of renormalizing the electron operators $c_a \to \sqrt{Z_a} c_a$.

This approach modifies the pairing interaction in orbital space and thus also influences the superconducting gap found when solving the linearized gap equation (*12,30*) or the Bogoliubov de Gennes equation self-consistently (*20,21*). In a simple picture, it suppresses pair scattering from and to $d_{xy}$ orbitals and $d_{xz}$ orbitals, as they are less coherent. Within this approach, the resulting gap gets much more anisotropic because the only left over orbital channel (with significant orbital contribution close to the Fermi level) is $d_{yz}$. To illustrate the drastic changes upon orbital selectivity, we strongly suppress the weight of the $d_{xy}$ orbital and moderately suppress the $d_{xz}$ orbital to calculate the spin susceptibility using Eq. (S6) instead of the bare Green function $G^0_{ab}(\mathbf{k},\omega)$. Calculating the pairing interaction and solving the linearized gap equation (*12,30*) we obtain a gap function as shown in Fig S18B. Obviously, the trends in the anisotropy and the relative magnitudes of the gaps are drastically changed. At the same time, the susceptibility becomes more

($\pi$, 0) dominated, consistent with Refs. (*27,28*). Finally, we note that the electronic states themselves may show properties of decoherence according to the quasiparticle weights in band space. To demonstrate the implementation of this effect, we perform a pairing calculation where additionally pairing in the $d_{xy}$ ($d_{xz}$) states is relatively suppressed according to the decoherence of these states described by the *Z*-factors. In this calculation, the decoherence enters via the susceptibility and as additional prefactors through the matrix elements when projecting the pairing interaction from orbital to band space, such that an almost perfect agreement between theoretical result and experiment is obtained, see Figs. S18C, S18D.

## IX. BAND STRUCTURE PARAMETERS

Hoppings $t^{ab}_{r-r'}$, used for the calculations, are attached in a separate file in a machine readable form. The format is the following: $r_x, r_y, r_z, a, b, Re(t), Im(t)$. The Hamiltonian as described above can be obtained by a Fourier transform of the following lines, e.g. the contribution of each line to the element $a$ row and $b$ column of the Hamiltonian $H_0$ is given by $exp\left(-i(r_xk_x + r_yk_y + r_zk_z)\right)[Re(t) + iIm(t)]$. Note that the basis of the orbitals has been chosen as in (*10*) yielding also complex hoppings. Note further that the orbital order term as well as the important contribution of the spin orbit coupling from $S_zL_z$ is also included in these hoppings.

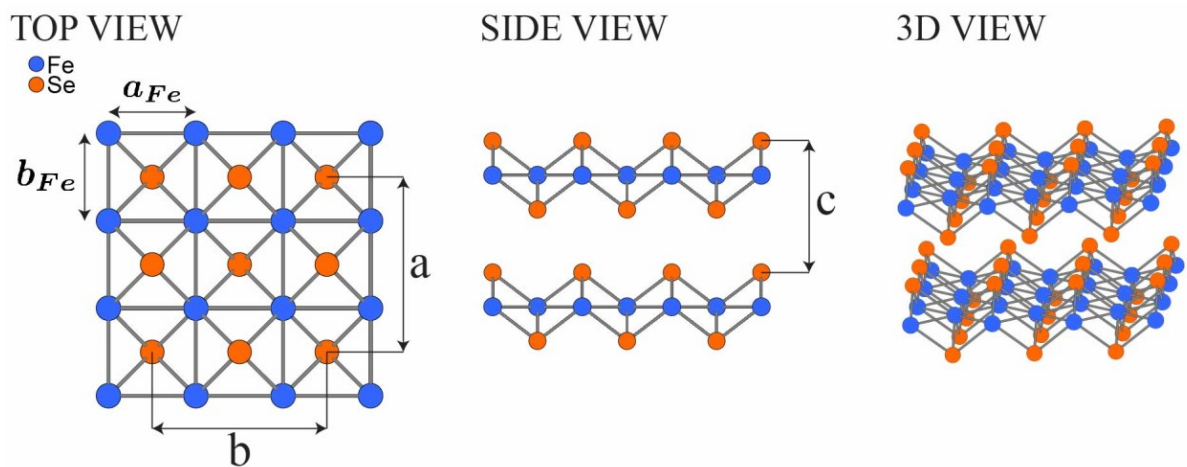

**Figure S1 | FeSe crystal structure.** The lattice parameters a = 5.31Å < b = 5.33Å < c = 5.48Å define the orthorhombic unit cell below the structural transition. In accordance with earlier work (Ref. 8,9) we introduce the non-standard parameters $a_{Fe} > b_{Fe}$ for labeling throughout this paper.

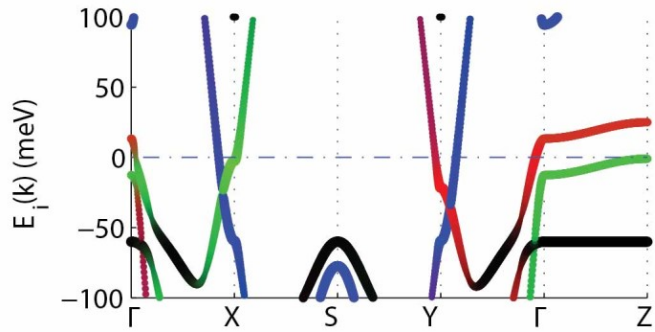
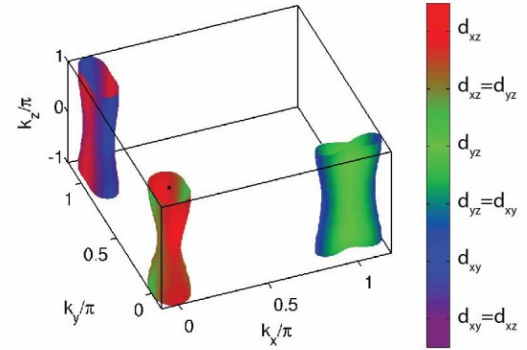

**Figure S2 | Tight-binding model for FeSe. A** Band structure used as a basis for theoretical calculations. Line thickness represents the magnitude of the dominant orbital content. Thus, if the line gets thinner orbital content will be more mixed. Red = $d_{xz}$, green = $d_{yz}$, blue = $d_{xy}$, and black = $d_{x^2-y^2}$ or $d_{z^2}$. **B** Fermi surface including orbital character of our band structure model at low temperatures.

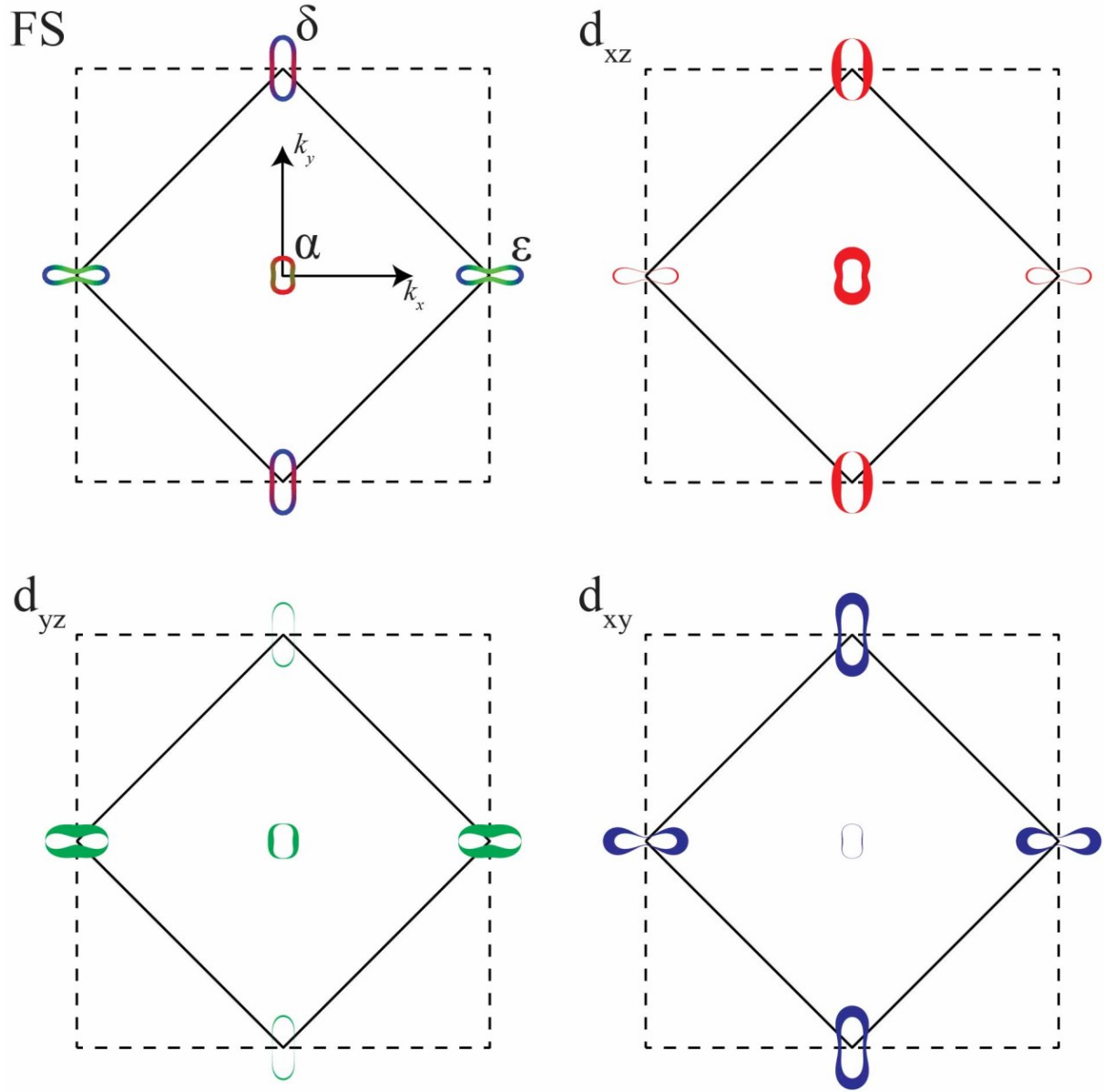

**Figure S3** | Upper left panel shows the Fermi surface at $k_z = 0$ for our band structure model (also shown figure 1B of the main manuscript). In the remaining three panels, the individual contributions of the main *d*-orbitals are shown explicitly on the Fermi surface (greater line thickness corresponding to bigger contribution).

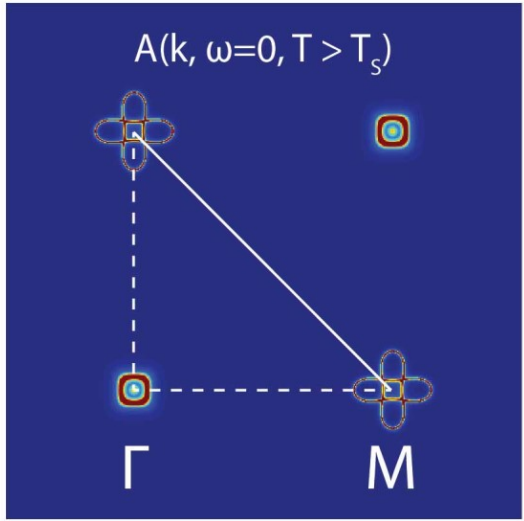 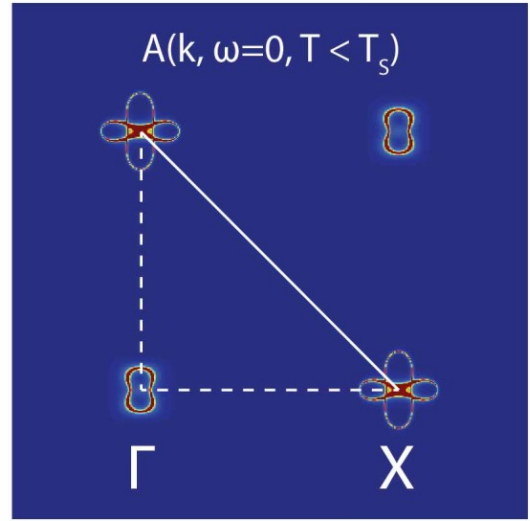

**Figure S4 | 10-band spectral function of the tight-binding model above and below the structural transition. A,** Above the structural transition the orbital order terms are zero. As a consequence of this, both the hole-like pockets around $\Gamma$ and the electron-like pockets around the M-point are $C_4$-symmetric. **B,** Below the structural transition, the orbital order terms break the $C_4$-symmetry of both the hole-like pocket around $\Gamma$ and the electron-like pockets around X.
The solid white line marks the boundary of the 2Fe-unit cell Brillouin zone, and the dashed white lines represent symmetry axes about which the Fermi surface is mirror symmetric.

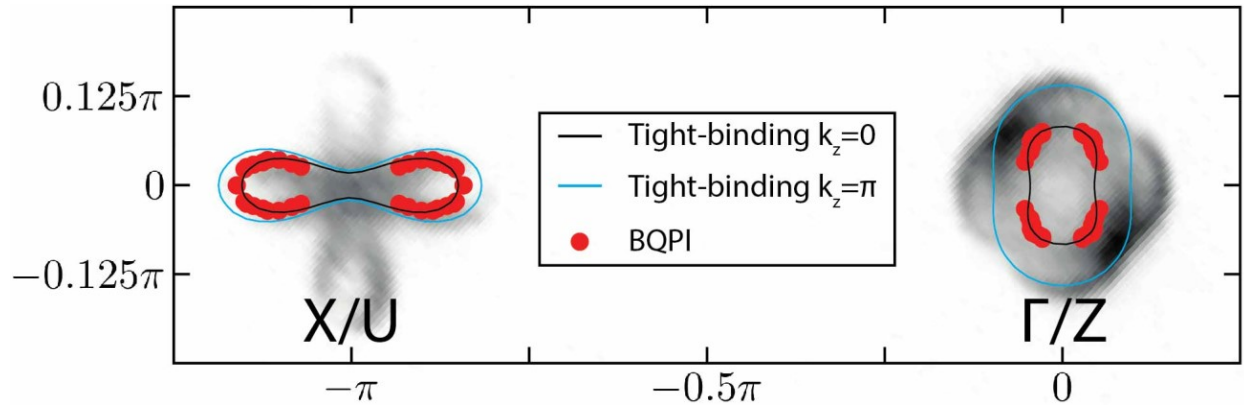

**Figure S5** | Comparison between the Fermi surface of our model at $k_z = 0$ (black lines) and $k_z = \pi$ (blue lines) and experimentally deduced points of the Fermi surface from BQPI (red dots) and a map of the spectral function at $\omega = 0$ and $k_z = \pi$ as measured by APRES (14) (gray map). The features in the ARPES spectral function that generate an apparent $C_4$-symmetry about both the Γ-point and about the X-point, are due to summation over the two types of nematic/orthorhombic domains that are orthogonal to each other, and are irrelevant for band structure parameterization.

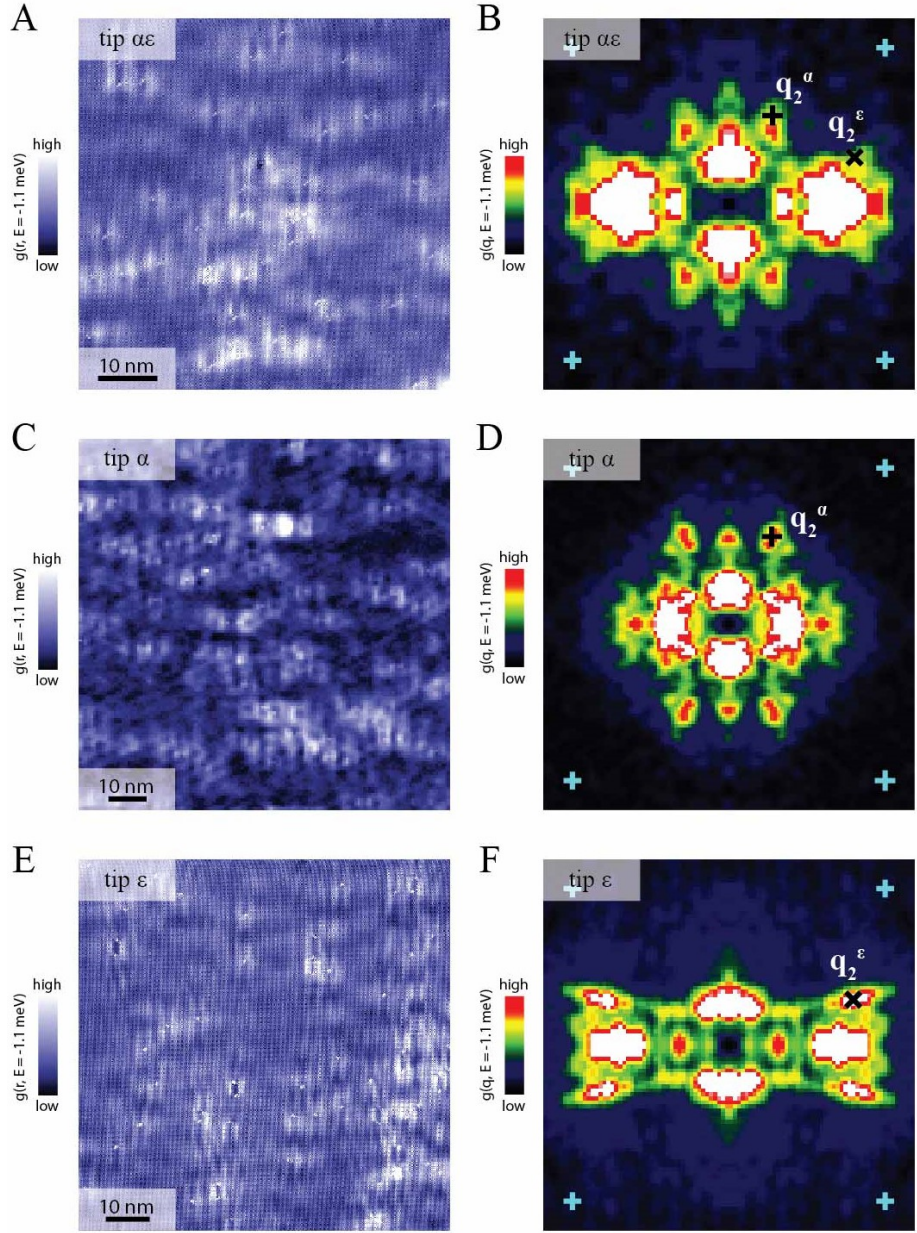

**Figure S6 | Different tunneling tips. A, C, E** Differential tunneling conductance images $g(\vec{r}, E)$ for three different tips at -1.1 meV. **B, D, F** Symmetrized, averaged, and core subtracted amplitude Fourier transforms $|g(\vec{q}, E)|$ of the conductance maps in A, C, E. The tip in A and B is simultaneously sensitive to both α- and ε-band BQPI. The tip in C and D is sensitive predominantly to the α pocket because it spatial resolution is low and so can only detect long wavelength BQPI. The tip in E and F instead mostly displays sensitivity to the ε-pocket. Cyan crosses mark $(\pm \frac{2\pi}{8a_{Fe}}, \pm \frac{2\pi}{8b_{Fe}})$ positions.

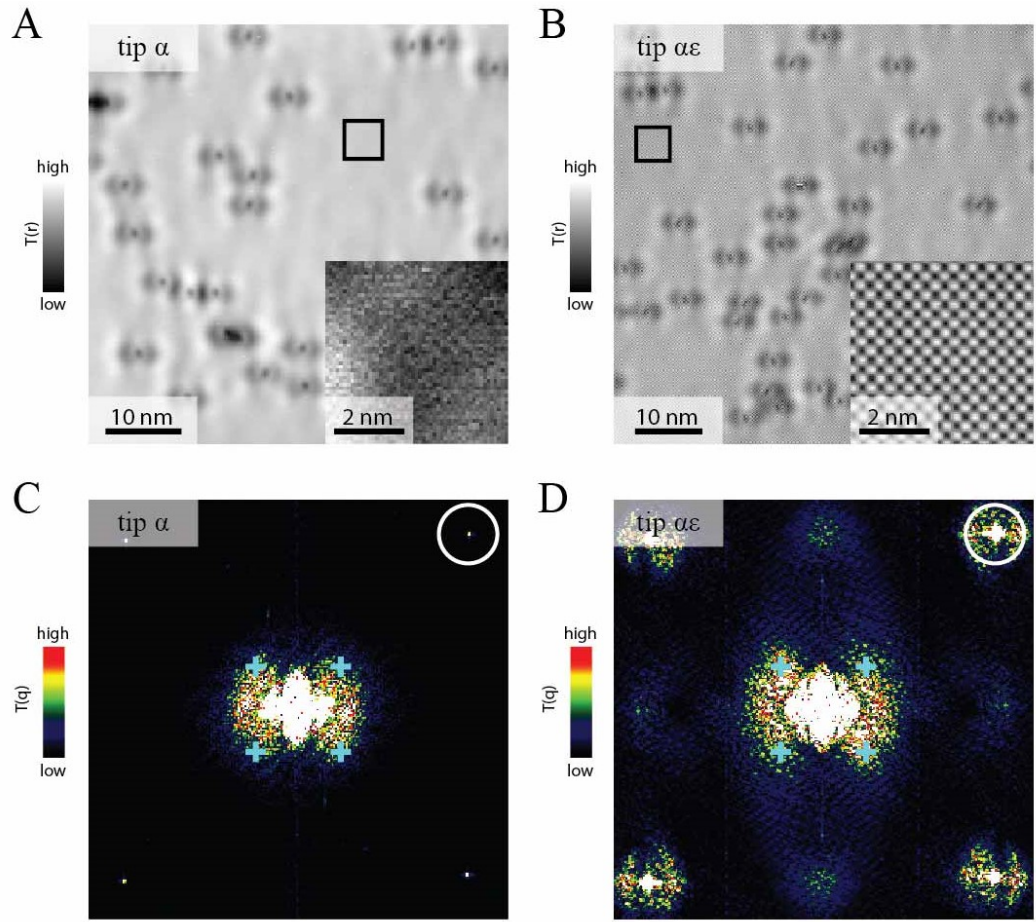

**Figure S7 | Relationship between r- and q-space sensitivity of different tunneling tips. A, B** Constant current topography in the same nematic domain in FeSe with two of the tips discussed in additional detail in the text and Fig. S6. The inset shows the topographic region marked by the black box. The ability to resolve atoms differs strongly between the two tips. **C, D** Amplitude of the Fourier transform of the topographs in A and B. The tip with superior spatial resolution detects scattering at high q-values, not observed in the Fourier transform of the lower spatial resolution tip. The white circle marks $(+\frac{\pi}{a_{Fe}}, +\frac{\pi}{b_{Fe}})$. Cyan crosses mark $(\pm\frac{2\pi}{8a_{Fe}}, \pm\frac{2\pi}{8b_{Fe}})$ positions.

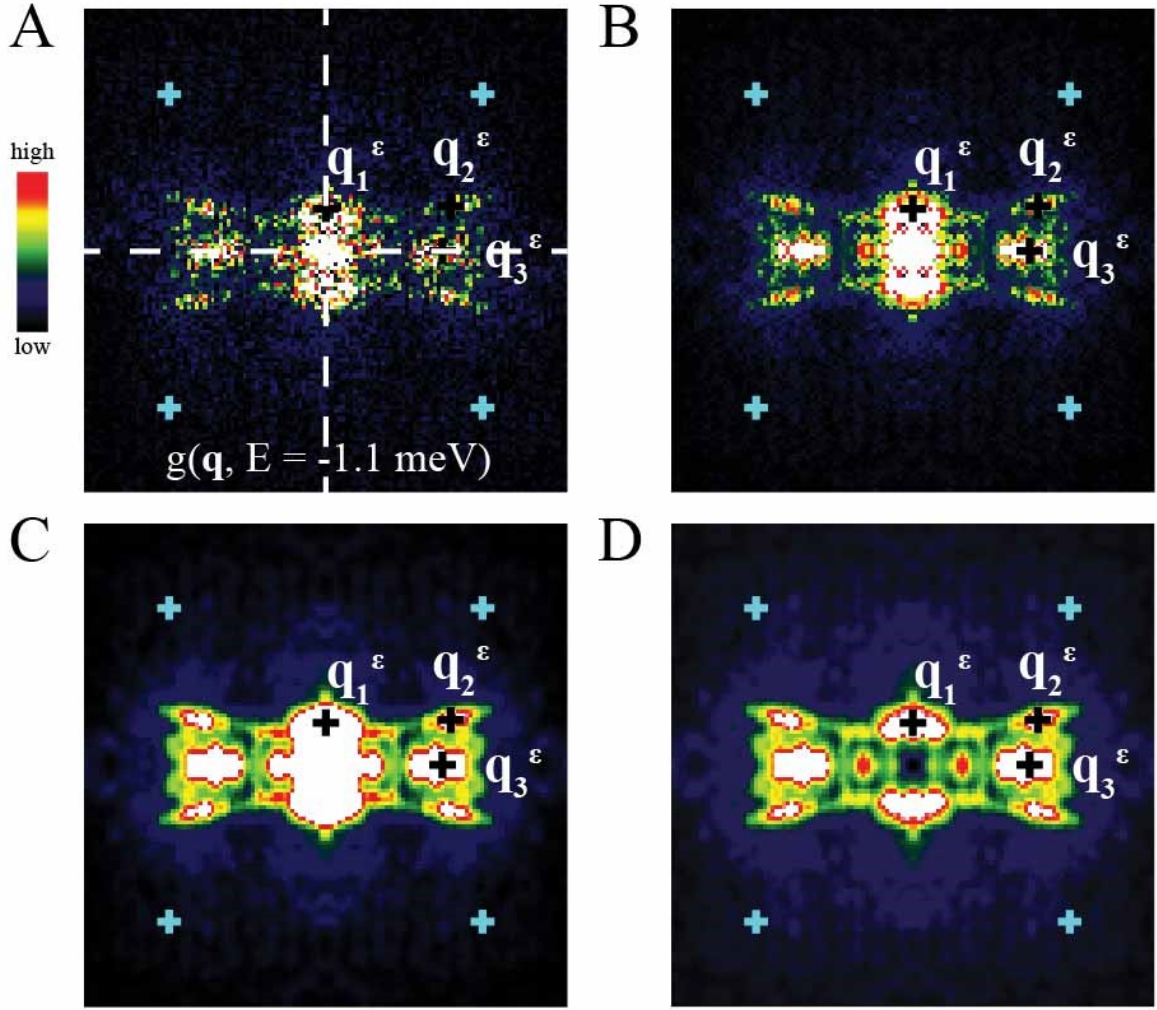

**Figure S8 | BQPI data processing steps. A,** Raw amplitude Fourier transform $|g(\vec{q}, E)|$ as obtained from the measured differential conductance shown in Fig. S6E. The dashed white lines represent mirror symmetry axes of the Brillouin zone, see also Fig. S4. **B,** Symmetrized amplitude Fourier transform $|g(\vec{q}, E)|$ created by reflection about the mirror symmetry axes displayed in A. **C,** Symmetrized and averaged amplitude Fourier transform $|g(\vec{q}, E)|$; after symmetrization a three-by-three pixel averaging filter is utilized in order to further increase signal-to-noise. **D,** Symmetrized, averaged and core subtracted amplitude Fourier transform $|g(\vec{q}, E)|$; in the last step a Gaussian core is subtracted in Fourier space which corresponds to a long wavelength filter in real space. In all panels cyan crosses mark $(\pm \frac{2\pi}{8a_{Fe}}, \pm \frac{2\pi}{8b_{Fe}})$ positions.

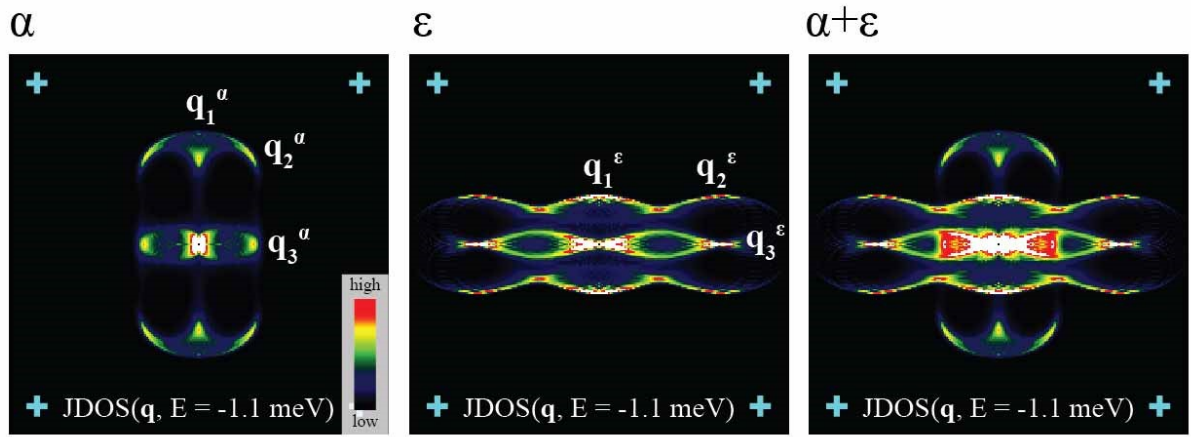

**Figure S9 | Partial Joint Density of States (JDOS)** using the tight-binding model and orbital selective pairing for the α-pocket, the ε-pocket and the sum of both the α- and ε-pocket. Cyan crosses mark $(\pm\frac{2\pi}{8a_{Fe}}, \pm\frac{2\pi}{8b_{Fe}})$ positions.

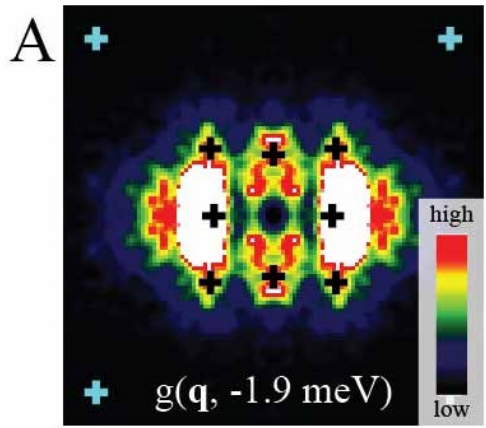
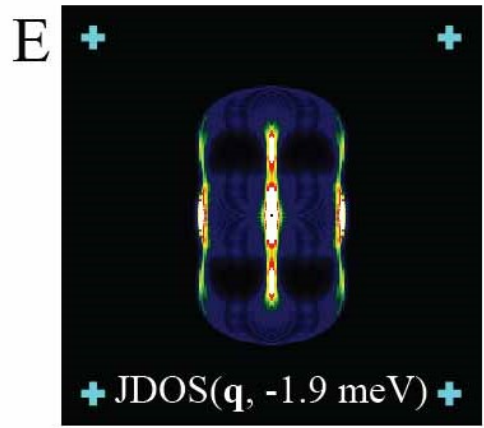

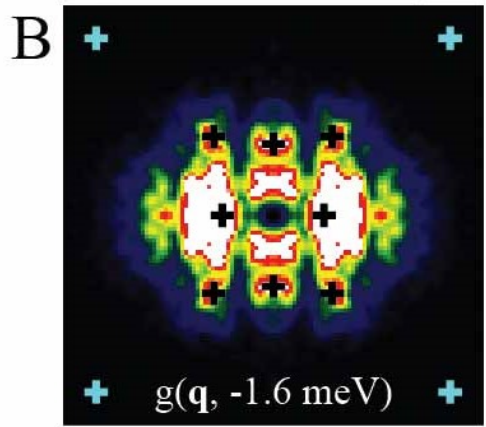
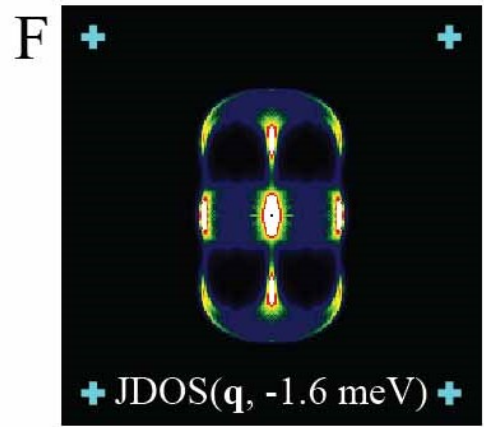

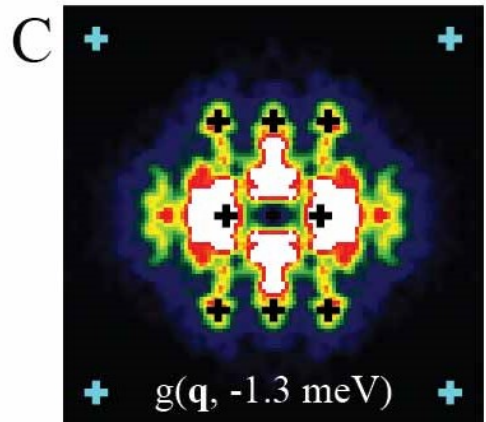
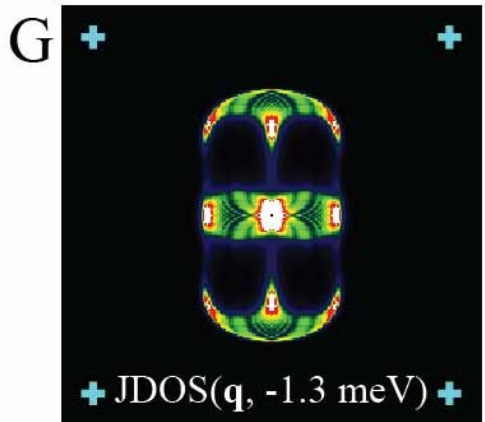

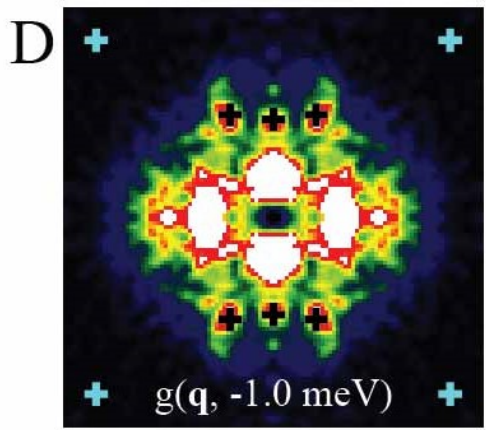
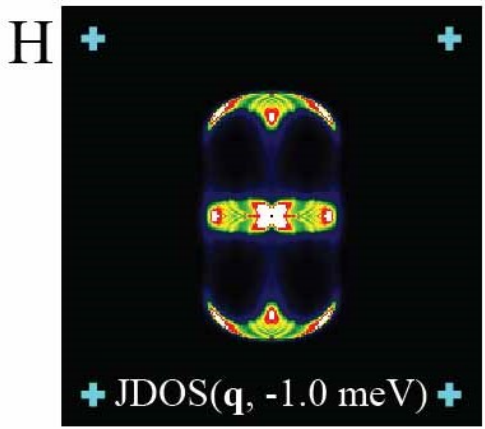

**Figure S10 | Comparison of $|g(q,E)|$ and partial JDOS for an $\alpha$-tip configuration**. **A - D**, Symmetrized, averaged, and core subtracted amplitude Fourier transforms $|g(\vec{q},E)|$ at four energies inside the superconducting gap $\Delta_\alpha$. Black crosses mark the extracted q-vectors expected from the 'banana' tips model. **E-H**, Partial JDOS for α-pocket at corresponding energies. Cyan crosses mark $(\pm\frac{2\pi}{8a_{Fe}}, \pm\frac{2\pi}{8b_{Fe}})$ positions.

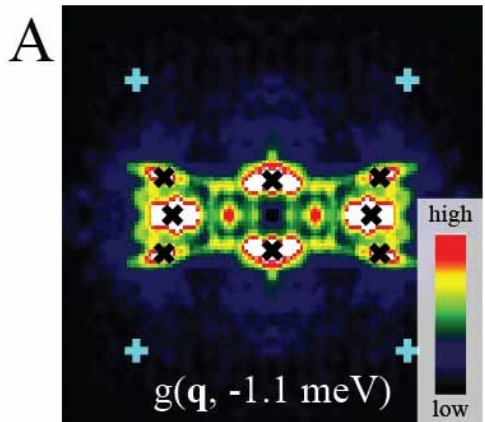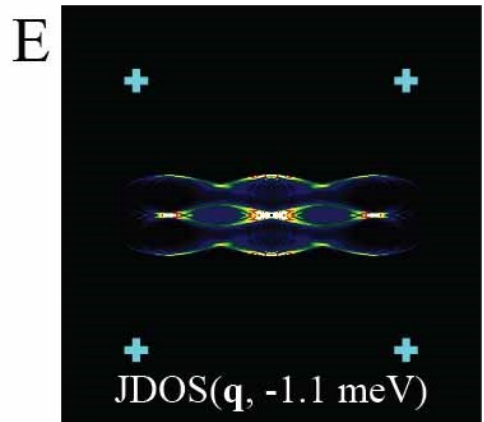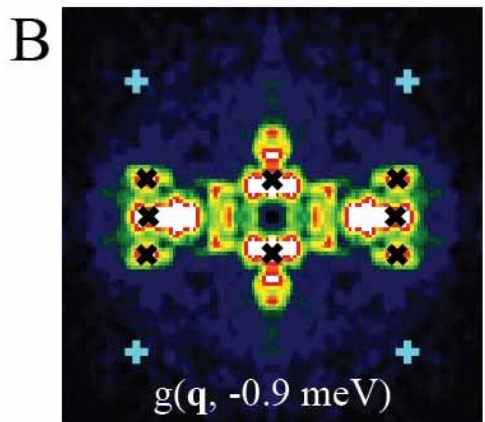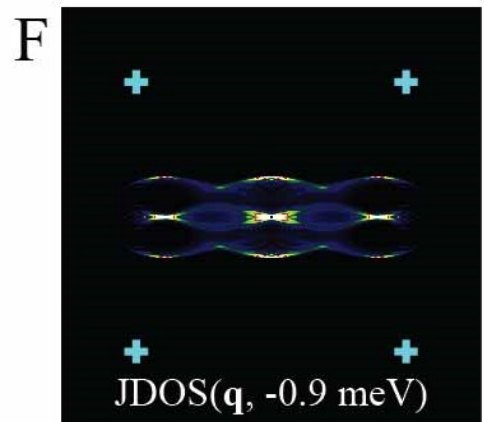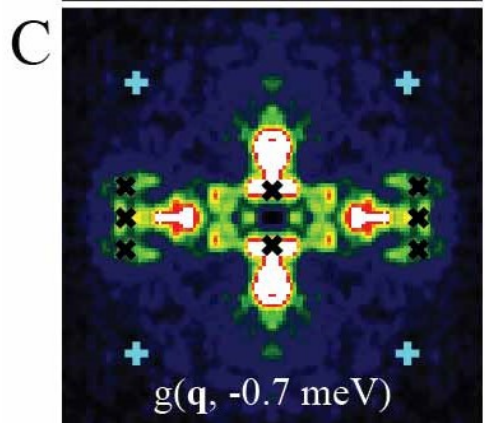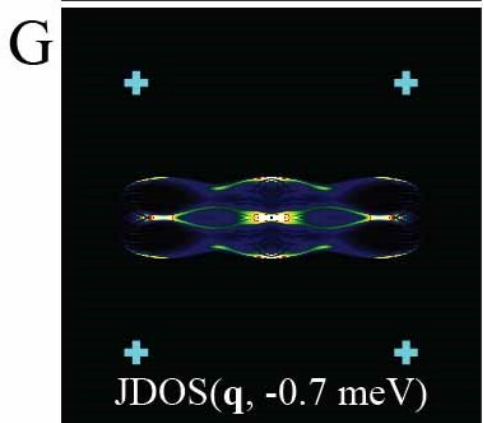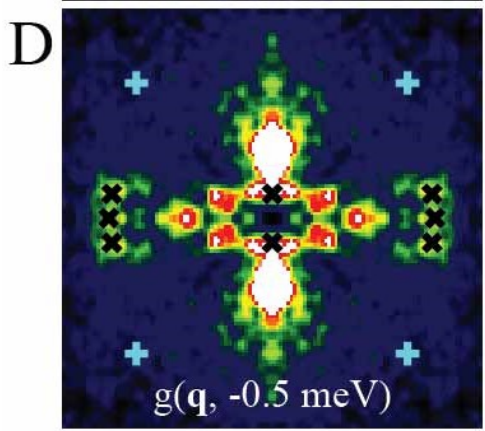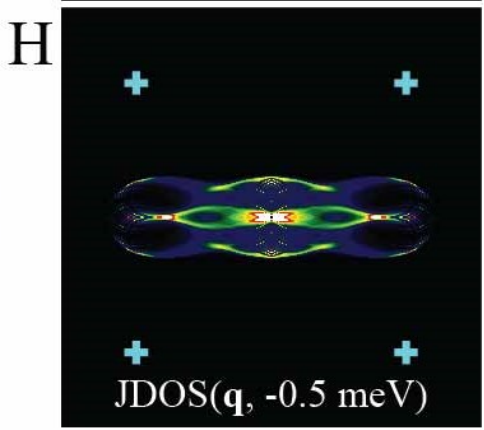

**Figure S11 | Comparison of $|g(q, E)|$ and partial JDOS for an $\varepsilon$-tip configuration**. **A - D**, Symmetrized, averaged, and core subtracted amplitude Fourier transforms $|g(\vec{q}, E)|$ at four energies inside the superconducting gap $\Delta_\varepsilon$. Black crosses mark the extracted q-vectors expected from the 'banana' tips model. **E-H**, Partial JDOS for $\varepsilon$-pocket at corresponding energies. Cyan crosses mark $(\pm \frac{2\pi}{8a_{Fe}}, \pm \frac{2\pi}{8b_{Fe}})$ positions.

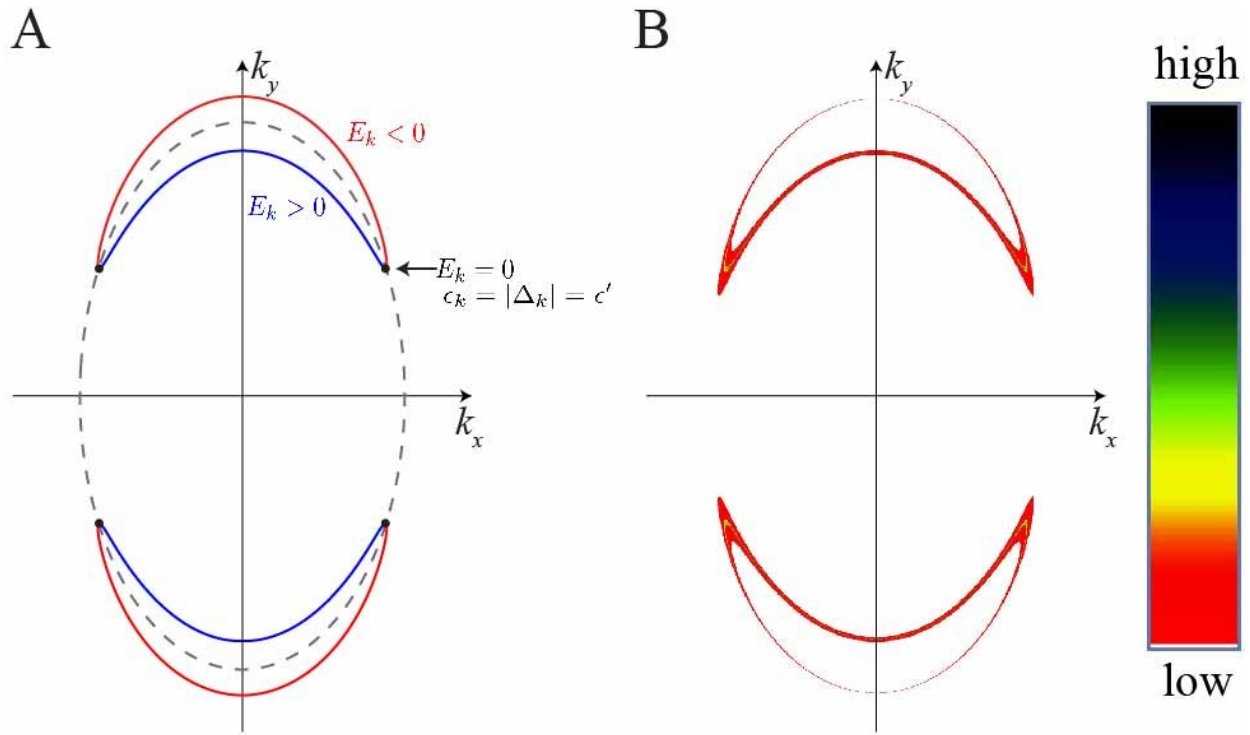

**Figure S12** | **A** Schematic showing a superconducting CEC at a particular energy $\epsilon' > 0$ (above the chemical potential) for an ellipsoidal hole pocket (grey dashed line) with two-fold symmetric anisotropic gap that is maximum along the $k_x$ directions and minimum along the $k_y$ direction. **B** Spectral function $A(\mathbf{k}, \omega = \epsilon')$ corresponding to A. The spectral weight is greatest at the locations of 'banana' tips.

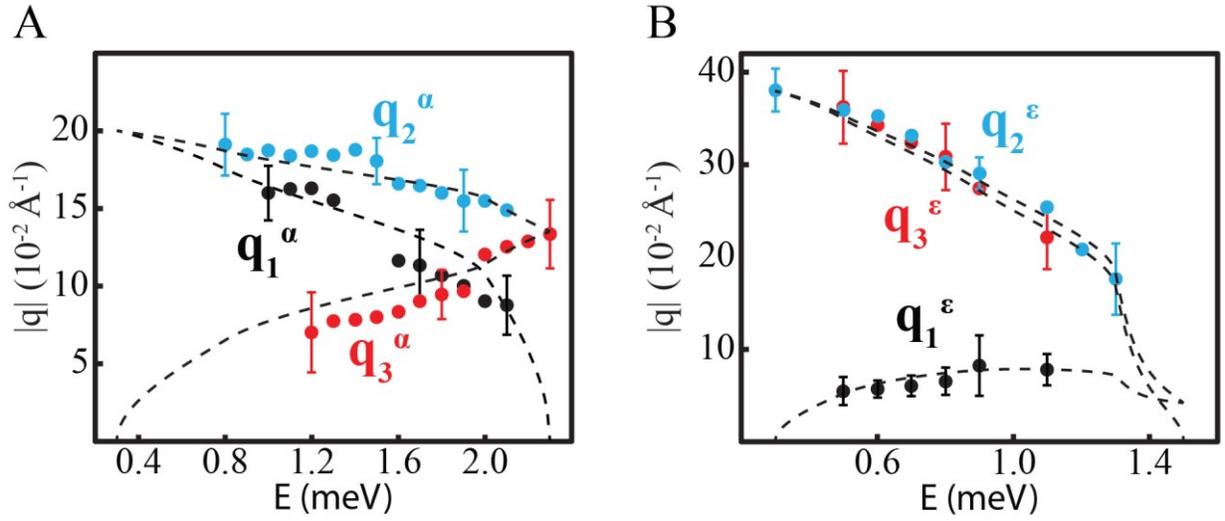

**Figure S13** | **A** Measured evolution of $|\vec{q}_i^{\alpha}(E)|$ for 2.3meV>|E|>0.8meV at 280mK. Dashed lines symbolize the expected energy dependence of $|\vec{q}_i^{\alpha}(E)|$ in a 'banana tips' model for empirically determined Fermi surface and $\Delta_{\alpha}(\vec{k})$. **B** Measured evolution of $|\vec{q}_i^{\varepsilon}(E)|$ for 1.3meV>|E|>0.3meV at 280mK. Dashed lines symbolize the expected energy dependence of $|\vec{q}_i^{\varepsilon}(E)|$ in a 'banana tips' model for empirically determined Fermi surface and $\Delta_{\varepsilon}(\vec{k})$.

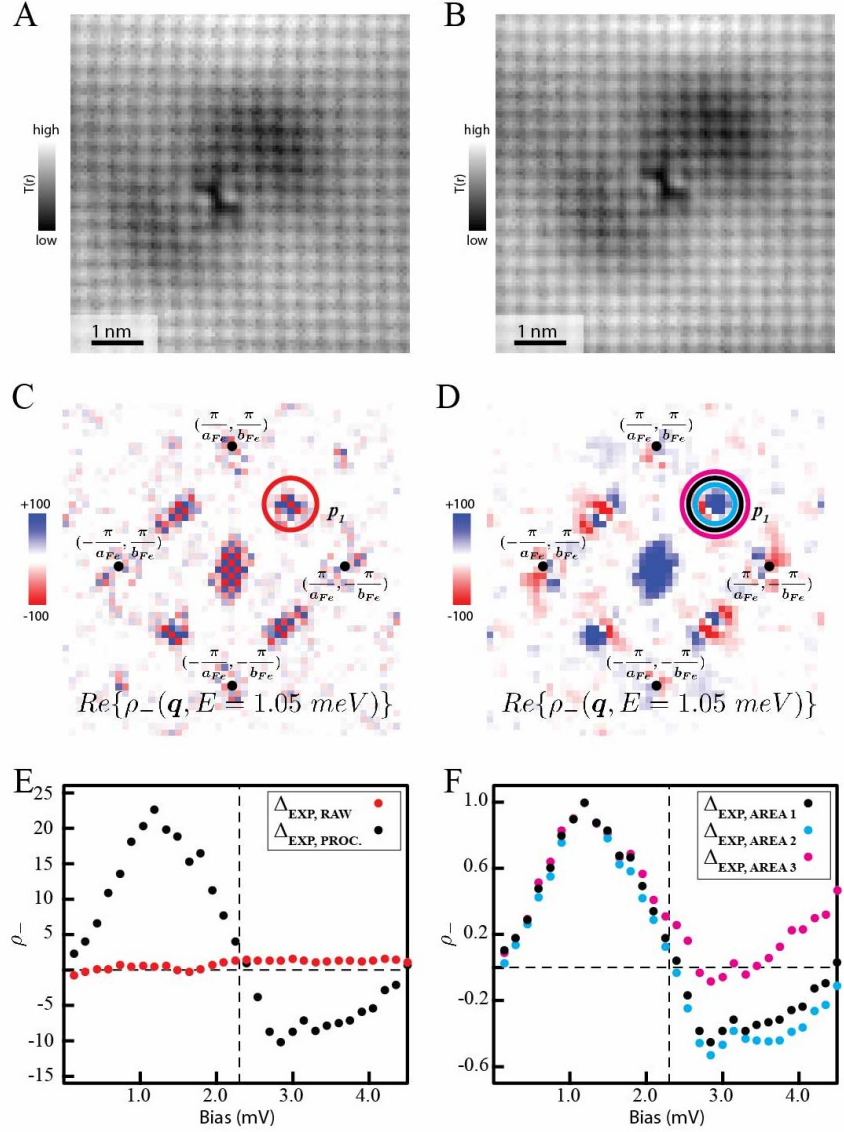

**Figure S14 | Experimentally extracted $\rho_-(\omega)$. A** Constant current topography of a single defect in FeSe. **B** Constant current topography of the same defect as in A, but the data has both been LF-corrected and shifted so that the center of the defect is at the origin of the FT. **C** $Re\{\rho_-(\vec{q}, \omega = 1.05\ meV)\}$ of the raw data. **D** $Re\{\rho_-(\vec{q}, \omega = 1.05\ meV)\}$ of the LF- and shift-corrected data. **E** Comparison of $\rho_-(\omega)$ for both the raw and LF- and shift-corrected data. **F** Comparison of $\rho_-(\omega)$ for three differently sized integration areas in $\vec{q}$-space. The vertical dashed black line marks the energy of the maximum superconducting gap in E and F.

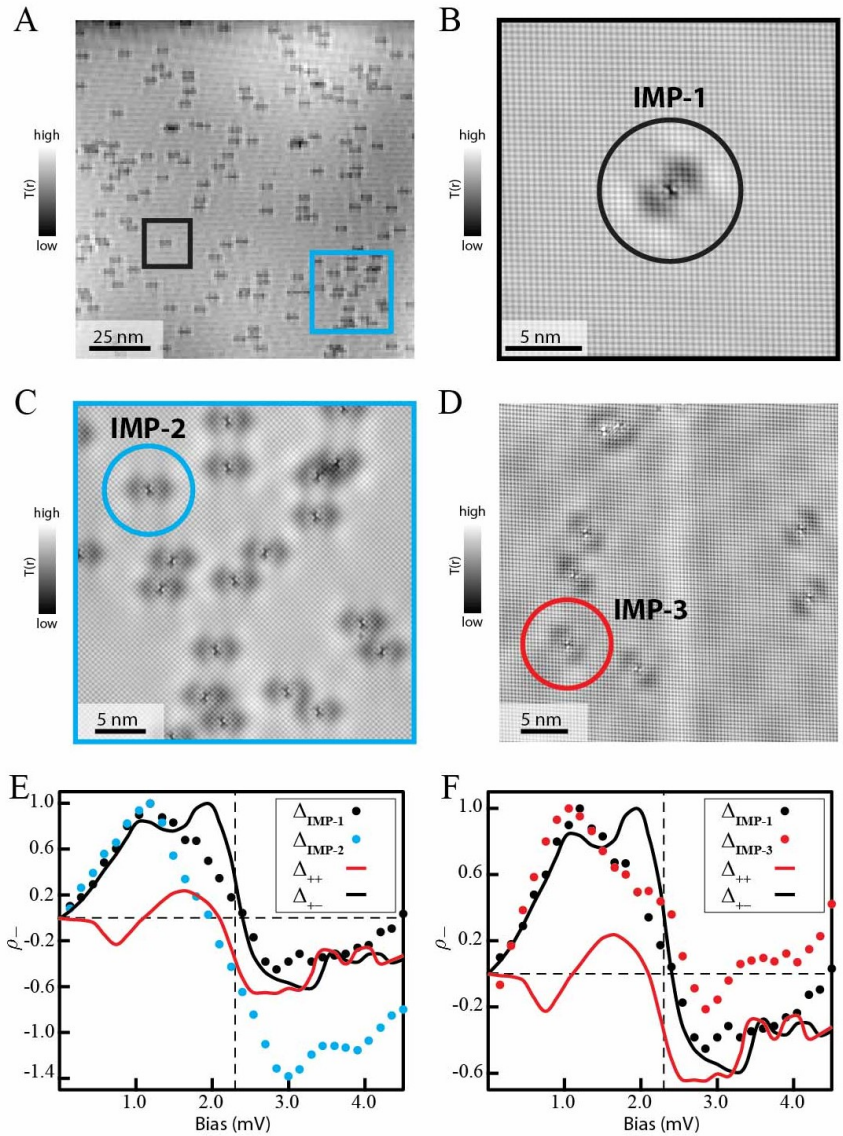

**Figure S15 | $\rho_-(\omega)$ for three different impurities.** **A** Constant current topography of a single domain in FeSe. The black and blue square mark the positions of the topograph presented in B and C, respectively. **B** Constant current topography of a single defect, referred to as impurity 1 in E and F. **C** Constant current topography of a several defects. The blue circle marks impurity 2. **D** Constant current topography of several defects near a twin boundary in a second sample of FeSe studied by SI-STM. The red circle marks impurity 3. **E** Comparison of $\rho_-(\omega)$ for impurities 1 and 2 of the same sample with theoretically predicted $\rho_-(\omega)$. **F** Comparison of $\rho_-(\omega)$ for impurities 1 and 3 of two different FeSe samples with theoretically predicted $\rho_-(\omega)$. The vertical dashed black line marks the energy of the maximum superconducting gap in E and F.

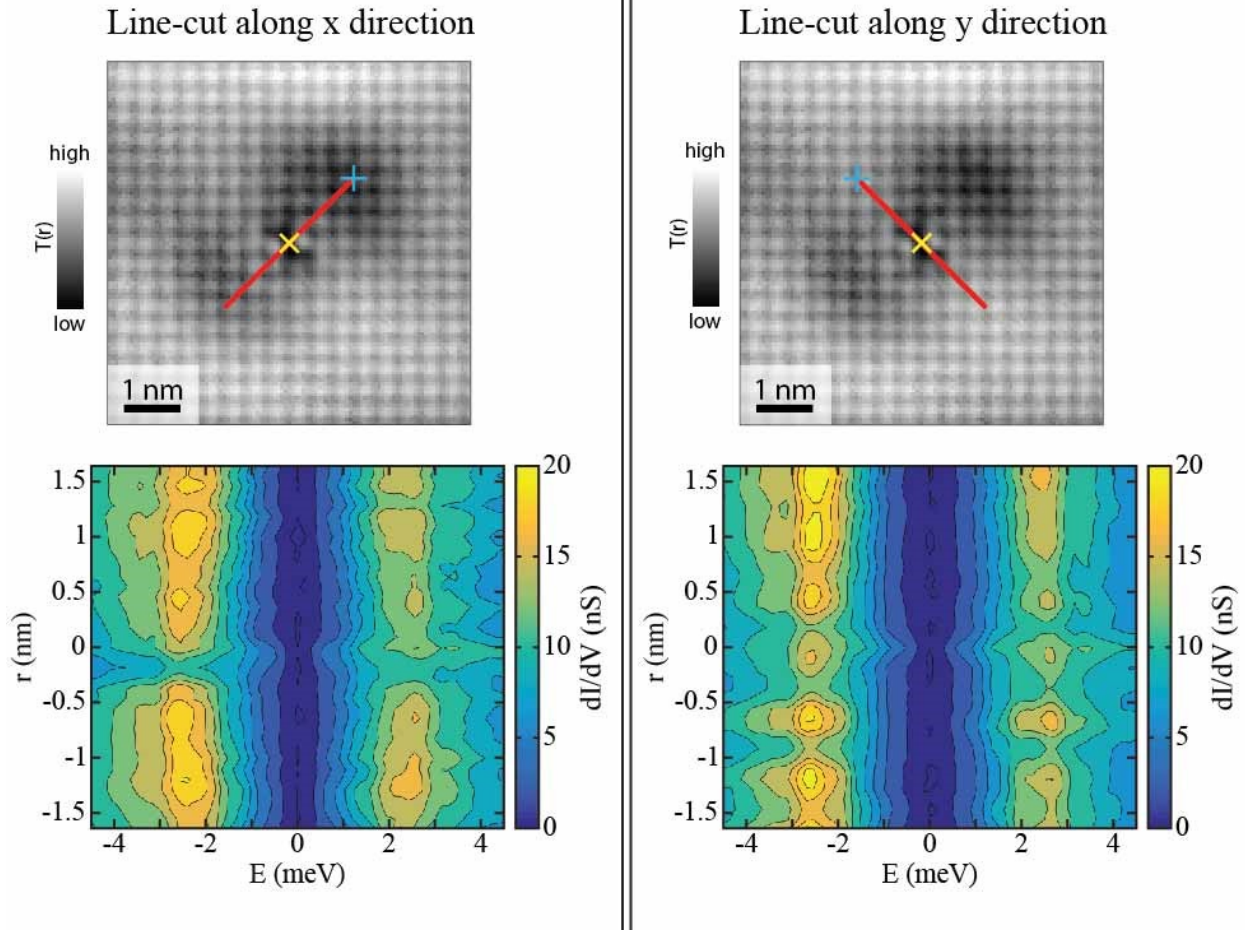

**Figure S16 | In-gap states induced by impurity.** dI/dV line-cuts through the single impurity site along the x- and y-direction where x is parallel to $a_{Fe}$ and y is parallel to $b_{Fe}$. While small, an increase of the dI/dV is clearly visible inside the superconducting gap at the location of the impurity (corresponding to 0 nm in the dI/dV panels) that is indicative of a weak in-gap impurity-state.

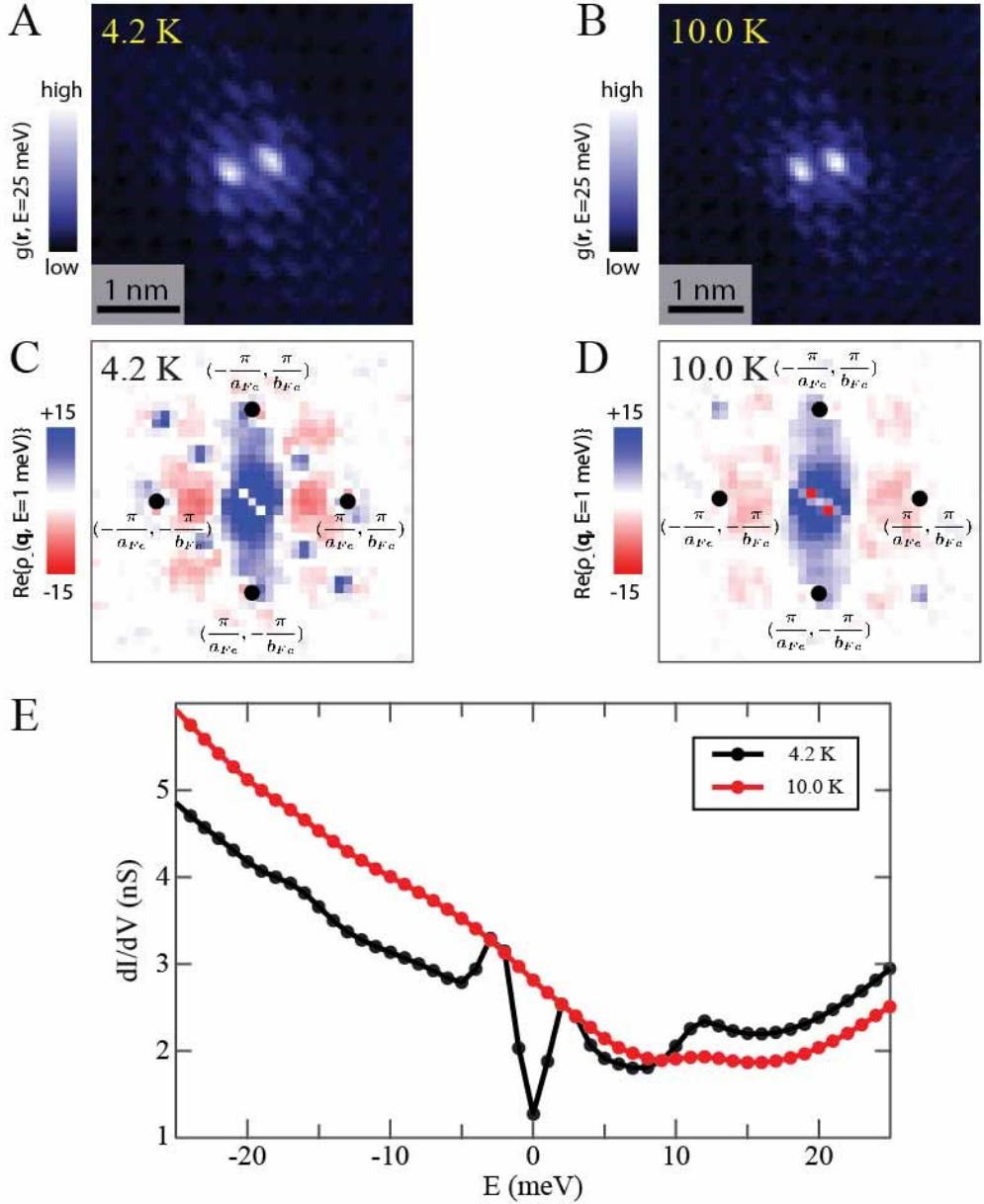

**Figure S17** | Disappearance of $\rho_-(\vec{q}, \omega)$-signal above $T_C$. **A, B,** Differential conductance image at 25 meV of the same impurity at 4.2 K and 10.0 K. **C, D,** $Re\{\rho_-(\vec{q}, \omega = 1\ meV)\}$ for the defect shown in A and B. The signal corresponding to scattering between the α- and ε-pocket vanishes above $T_C$. **E,** Average spectrum for the field of view shown in A, B recorded for the same setup voltage and current. Above $T_C$ no sign of the superconducting gap remains.

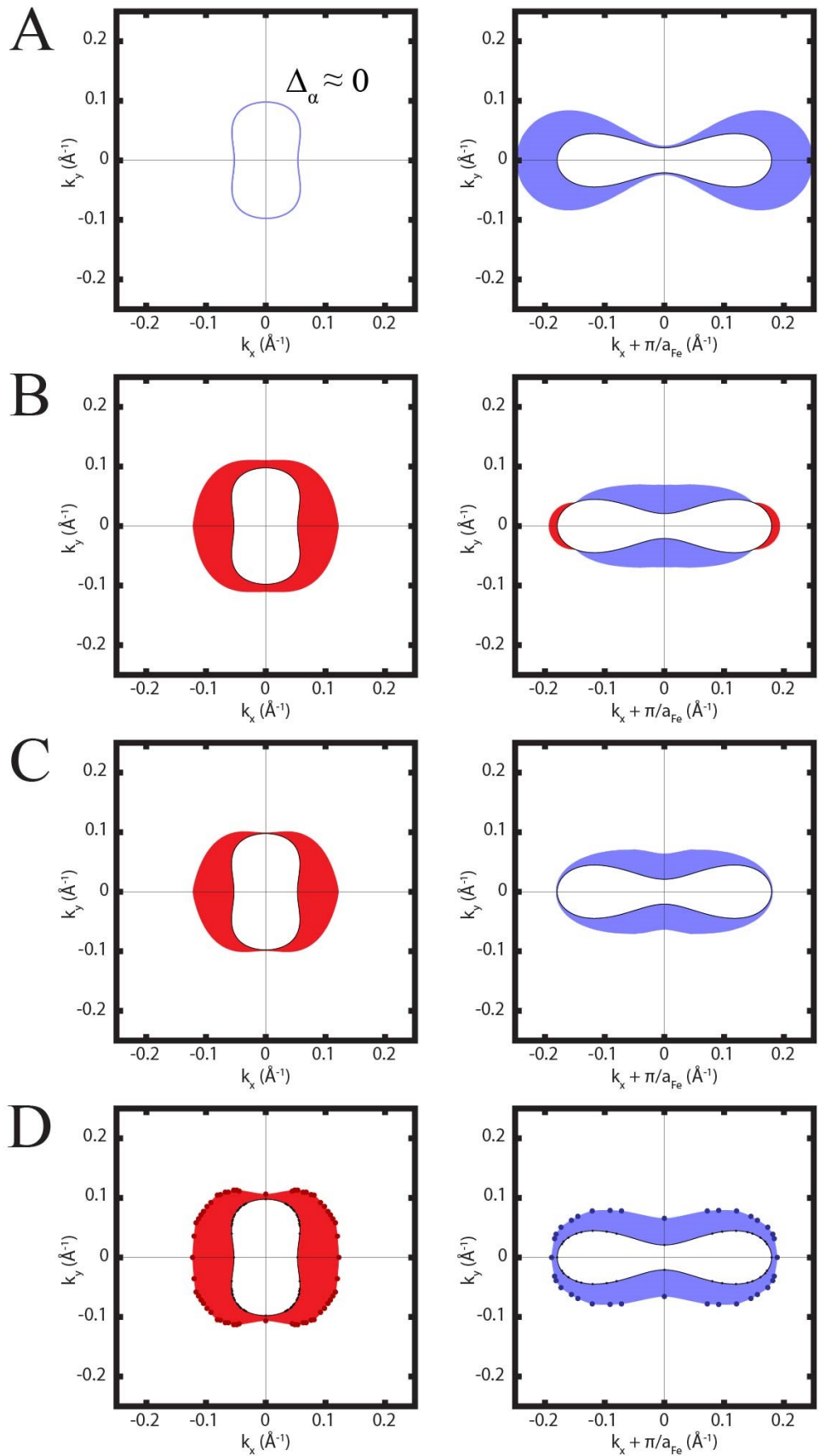

**Figure S18 | Results from calculations of the gap-symmetry function** (*12,30*). **A**, traditional spin-fluctuation theory of the present model, yielding a very small gap on the $\alpha$ pocket and a large anisotropic gap on the $\varepsilon$ (and $\delta$ pocket; not shown). Comparing to the experimental result [panel D] shows a strong discrepancy qualitatively and quantitatively. **B**, spin fluctuation pairing using a susceptibility calculated with modified quasiparticle weights $Z_\alpha$, Eq. (S6), yields a gap function that has some similarities to the measured order parameter from BQPI, whereas a nodal feature on the $\varepsilon$ band together with the sign-change disagrees with experimental findings. **C**, additionally imposing orbital selectivity in the pairing itself, e.g. suppressing pairing in the $d_{xy}$ channel and in the $d_{xz}$ channel according to their respective decoherence as described by the Z-factors when projecting the pairing interaction to momentum space, yields almost perfect agreement due to the dominant $d_{yz}$ pairing. **D**, superconducting gap deduced experimentally from BQPI. The gap symmetry functions in all models (A-C) were scaled to have the same maximum gap value that would agree with the main coherence peaks observed in dI/dV spectra.

**Movie S1**

Movie contains three panels showing measured and simulated BQPI for the α-pocket. From left to right the content is as follows: $g(\vec{r}, E)$; symmetrized, averaged, and core subtracted $|g(\vec{q}, E)|$ where the black crosses mark the position of $\vec{q}_i^{\alpha}(E)$ i=1-3 and blue crosses mark the $(\pm\frac{2\pi}{8a_{Fe}}, \pm\frac{2\pi}{8b_{Fe}})$ positions; partial $JDOS(\vec{q}, E)$ for the α-pocket where the blue crosses mark the $(\pm\frac{2\pi}{8a_{Fe}}, \pm\frac{2\pi}{8b_{Fe}})$ positions.

**Movie S2**

Movie contains three panels showing measured and simulated BQPI for the ε-pocket. From left to right the content is as follows: $g(\vec{r}, E)$; symmetrized, averaged, and core subtracted $|g(\vec{q}, E)|$ where the black crosses mark the position of $\vec{q}_i^{\varepsilon}(E)$ i=1-3 and blue crosses mark the $(\pm\frac{2\pi}{8a_{Fe}}, \pm\frac{2\pi}{8b_{Fe}})$ positions; partial $JDOS(\vec{q}, E)$ for the ε-pocket where the blue crosses mark the $(\pm\frac{2\pi}{8a_{Fe}}, \pm\frac{2\pi}{8b_{Fe}})$ positions.

**Movie S3**

$\rho_-(E)$ and $\rho_-(\vec{q}, E)$ for single, isolated impurity in FeSe. The black circle surrounds the area in $\vec{q}$-space for interband scattering $\vec{p}_1$ between the α- and ε-pocket.

**Additional Data table S1 (separate file)**

Hoppings $t_{r-r'}^{ab}$ of the tight-binding model used for the calculations, are attached in a separate file in a machine readable form. The name of the file is 'tb_FeSe_parameters.csv'. Columns in the file are arranged as follows: $r_x, r_y, r_z, a, b, Re(t), Im(t)$. For more information, see section IX of the supplementary text.